\documentclass[twocolumn,twocolappendix]{aastex631}

\graphicspath{{./}{figures/}}

\usepackage{graphicx}
\usepackage{xcolor}
\usepackage{amsmath,bm}
\usepackage[T1]{fontenc}
\usepackage{rotating}
\usepackage{multirow}
\usepackage{hhline}

\newcommand{\cp}  {$C_\mathrm{P}$}
\newcommand{\dnds}{$\diff N/ \diff S$}

\newcommand{\diff}{\mathrm{d}}
\newcommand{\I}[1]{\textit{#1}}
\newcommand{\ie}  {\emph{i.e.}}
\newcommand{\de}{\mathrm d}

\newcommand{\Fermi}{{\sl Fermi}}

\newcommand{\sv}{\sigma_{\rm ann} v}
\newcommand{\mdm}{{m_{\rm DM}}}
\newcommand{\odm}{{\Omega_\mathrm{DM}}}

\newcommand{\mr}[1]{\mathrm{#1}}

\def \aap   {A\&A}

\def \apjs  {ApJS}
\def \apjl  {ApJL}

\begin{document}

\title{Flat spectrum radio quasars and BL Lacs dominate the anisotropy \\  of the unresolved gamma-ray background}
\shorttitle{fsrqs and blls dominate the anisotropy of the ugrb}
\shortauthors{Korsmeier et al.}

\author[0000-0003-3478-888X]{Michael Korsmeier}
\email{michael.korsmeier@fysik.su.se}
\affiliation{The Oskar Klein Centre, Department of Physics, Stockholm University, \\ AlbaNova, SE-10691 Stockholm, Sweden}
\affiliation{Dipartimento di Fisica, Universit\`a di Torino, \\Via P. Giuria 1, 10125 Torino, Italy}
\affiliation{Istituto Nazionale di Fisica Nucleare, Sezione di Torino, \\Via P. Giuria 1, 10125 Torino, Italy}
\affiliation{Institute for Theoretical Particle Physics and Cosmology, RWTH Aachen University, \\Sommerfeldstr.\ 16, 52056 Aachen, Germany}

\author[0000-0001-7070-0094]{Elena Pinetti}
\email{elena.pinetti@unito.it, elena.pinetti@to.infn.it}
\affiliation{Dipartimento di Fisica, Universit\`a di Torino, \\Via P. Giuria 1, 10125 Torino, Italy}
\affiliation{Istituto Nazionale di Fisica Nucleare, Sezione di Torino, \\Via P. Giuria 1, 10125 Torino, Italy}
\affiliation{Laboratoire de Physique Théorique et Hautes Energies, CNRS \& Sorbonne Université, \\4 place Jussieu, F-75252 Paris, France}
\affiliation{Theoretical Astrophysics Department, Fermi National Accelerator Laboratory, Batavia, Illinois, 60510, USA}

\author[0000-0002-6548-5622]{Michela Negro}
\email{mnegro1@umbc.edu}
\affiliation{University of Maryland, Baltimore County, Baltimore, MD 21250, USA}
\affiliation{NASA Goddard Space Flight Center, Greenbelt, MD 20771, USA}
\affiliation{Center for Research and Exploration in Space Science and Technology, NASA/GSFC, Greenbelt, MD 20771, USA}

\author[0000-0003-0399-0284]{Marco Regis}
\email{marco.regis@unito.it, marco.regis@to.infn.it}
\affiliation{Dipartimento di Fisica, Universit\`a di Torino, \\Via P. Giuria 1, 10125 Torino, Italy}
\affiliation{Istituto Nazionale di Fisica Nucleare, Sezione di Torino, \\Via P. Giuria 1, 10125 Torino, Italy}

\author[0000-0002-3074-3118]{Nicolao Fornengo}
\email{nicolao.fornengo@unito.it, nicolao.fornengo@to.infn.it}
\affiliation{Dipartimento di Fisica, Universit\`a di Torino, \\Via P. Giuria 1, 10125 Torino, Italy}
\affiliation{Istituto Nazionale di Fisica Nucleare, Sezione di Torino, \\Via P. Giuria 1, 10125 Torino, Italy}

\begin{abstract}
We analyze the angular power spectrum (APS) of the unresolved gamma-ray background (UGRB) emission and combine it with the measured properties of the resolved gamma-ray sources of the \Fermi-LAT 4FGL catalog. Our goals are to dissect the composition of the gamma-ray sky and to establish the relevance of different classes of source populations of active galactic nuclei in determining the observed size of the UGRB anisotropy, especially at low energies. We find that, under physical assumptions for the spectral energy dispersion, \emph{i.e.} by using the 4FGL catalog data as a prior, two populations are required to fit APS data, namely flat spectrum radio quasars (FSRQs) at low energies and BL Lacs (BLLs) at higher energies. The inferred luminosity functions agree well with the extrapolation of the FSRQ and BLL ones obtained from the 4FLG catalog. We use these luminosity functions to calculate the UGRB intensity from blazars, finding a contribution of 20\% at 1GeV and 30\% above 10 GeV. Finally, bounds on an additional gamma-ray emission due to annihilating dark matter are derived.
\end{abstract}

%===================================================================
\section{Introduction}
\label{sec::intro}
%===================================================================

The extragalactic gamma-ray sky has been surveyed by the \Fermi\ Large Area Telescope (LAT) since the summer of 2008 \citep{2009ApJ...697.1071A}. The outstanding capability of this instrument has been groundbreaking in several aspects of high energy astrophysics. One important result is the detection and cataloging of extragalactic gamma-ray sources. The 8-year \Fermi-LAT source catalog, called the 4FGL catalog \citep{Fermi-LAT:2019yla}%
\footnote{This catalog is now also called 4FGL-DR1.},
counts more than 3300 extragalactic sources, which is more than 60\% of the entire catalog. Almost the totality of the extragalactic sources are blazars, a sub-class of active galactic nuclei (AGNs) with a jet pointing towards us: 35\% are BL Lacs (BLL), about 22\% are flat spectrum radio quasar (FSRQs) and about 41\% are blazars of unknown type (BCU). 

On top of the numerous detected extragalactic sources, even more numerous sub-threshold sources populate the unresolved gamma-ray background (UGRB).%
\footnote{The UGRB is also called isotropic gamma-ray background (IGRB) in literature. While for intensity studies it can be considered as isotropic, to a deeper level it is definitely not. Since in this paper we study its anisotropies, it is more appropriate to call it \emph{unresolved} instead.}
The UGRB emission represents about 20\% of the total gamma-ray emission and offers a unique observable of the extragalactic gamma-ray sky below the \Fermi-LAT source detection threshold. 

The UGRB is by definition a mission-time dependent component: The more the \Fermi-LAT surveys the sky, the more sensitive it becomes to less bright sources, leaving only the faintest objects unresolved. Guaranteed contributors to the UGRB emission are sub-detection-threshold blazars \citep{PhysRevD.86.063004, DiMauro:2017ing}, misaligned AGNs (mAGNs) \citep{DiMauro2013}, and star-forming galaxies (SFG) \citep{MARoth2021, 2014JCAP...09..043T}. Additionally, we cannot exclude contributions from more exotic components, such as dark matter (DM) \citep{Ando:2009, Calore:2013yia, Ajello:2015mfa, Fornasa:2016ohl, Zechlin:2017uzo} 

The UGRB emission has been studied through three main observables: its energy spectrum \citep{osti_1022500, Ackermann_2015}, its 1-point probability distribution function (1pPDF) through the photon counts statistics \citep{2016ApJS..225...18Z, 2016ApJ...826L..31Z, 2016ApJ...832..117L, DiMauro:2017ing}, and its angular power spectrum (APS) \citep{2012PhRvD..85h3007A, Fornasa:2016ohl, Ackermann:2018wlo}. The latter two observables investigate fluctuations over the UGRB isotropic emission to infer the properties of the underlying sources at the sub-threshold level. In this unresolved regime, mAGNs and SFGs, fainter than blazars but extremely more numerous, are expected to dominate the UGRB energy spectrum \citep{DiMauro2013, MARoth2021}. At the same time, at the current level of sensitivity, the blazars produce a higher level of spatial anisotropy than mAGNs and SFGs, and hence the former are expected to dominate the APS of the UGRB \citep{Mauro_2014, PhysRevD.86.063004}. SFGs and mAGNs could eventually emerge once the majority of the blazars have been resolved. Moreover, an improvement of the sensitivity is necessary in order to reveal the large-scale structure (LSS) of the Universe traced by gamma-ray sources (see e.g. \cite{Ando:2009}) that is encoded in a multipole-dependent APS. It is, therefore, crucial to update the UGRB anisotropy measurement in parallel to the detection of more sources in the LAT catalogs. 

The latest UGRB anisotropy measurement has been performed by the \Fermi-LAT Collaboration in 2018 \citep{Ackermann:2018wlo}. In that work, 8 years of Pass-8 (R3) data were analyzed, and it was consistently adopted the source catalog based on the same amount of observation time (FL8Y, a preliminary version of the 4FGL). The APS of the UGRB was measured in 12 energy bins between 500 MeV and 1 TeV. Additionally, the cross-correlation signal between the different energy bins (generically denoted by $i$ and $j$) was derived. In all cases, the APS (above $\ell=50$) was compatible with a constant value, C$_P^{ij}$, with no hint of LSS signature in the multipole range considered. This result confirms that the UGRB intensity fluctuation field, at the current level of sensitivity of the detector to point sources is still dominated by a population of relatively bright and not very numerous sources, so that the isotropically distributed fluctuations from Poisson noise dominate over the correlation due to clustering. Additionally, the anisotropy energy spectrum revealed a preference for a double power-law trend (with a high energy exponential cutoff) over a single power law (with a high energy exponential cutoff), placing a spectral break around 5 GeV.

Previous interpretation works, based on antecedent measurements of the UGRB anisotropy energy spectrum, were devoted to determining the components that contribute to the measured signal. In particular, \citep{Ando:2017alx} studied the results of \citep{Fornasa:2016ohl} and inferred the presence of a second steeper component, in addition to the blazar-only model, emerging below 2 GeV. However, the very soft spectral index implied by this analysis challenges the interpretation in terms of a known source population. Recently, \citep{Manconi:2019ynl} combined the 1pPDF, using methods as in \citep{2016ApJ...826L..31Z}, and the latest measurement of the anisotropy energy spectrum of the UGRB by \citep{Ackermann:2018wlo} to test blazars models (yet not distinguishing between BLLs and FSRQs) as well as the source count distribution of blazars extracted from the 4FGL catalog. They found that the assumption of the UGRB fluctuation field being entirely dominated by blazars is in agreement with both observables, which appear to show remarkable complementarity. Past works focused also on the DM interpretation of the UGRB anisotropy, as \citep{Fornasa:2016ohl}, where numerical simulations were used to model the DM distribution and its uncertainty in order to constrain the contribution from weakly interacting massive particles (WIMPs) in Galactic and extragalactic structures. The derived bounds are in the same ball-park as for other UGRB probes, but still significantly above the so-called thermal WIMP scenario. For a comprehensive overview about UGRB-related measurements and interpretation works prior \citep{Fornasa:2016ohl}, we address the reader to the review of \citep{Fornasa:2015qua}. 

In this work, we will focus on the latest measurement of the UGRB anisotropy energy spectrum \citep{Ackermann:2018wlo}. We investigate the contributions of different blazar types, distinguishing between BLLs and FSRQs. We find that BLLs and FSRQs can account for the totality of the UGRB anisotropy and also well reproduce the spectral features observed by \citep{Ackermann:2018wlo}. The analysis allows us to constrain many of the most relevant parameters of the blazar models in the unresolved regime. In a second step, we include the contribution to the UGRB arising from an annihilating DM particle and perform a global analysis to derive constraints on the particle DM parameters.  We account for both Galactic and extragalactic DM contributions, under different assumptions of the DM subhalo contribution and by including cross-terms in the anisotropy APS, due to the cross-correlation of the blazars contribution with the DM halos hosting them. 

The paper is structured as follows: Section~\ref{sec::blz} is devoted to blazars and we describe the blazar model adopted in our study, we introduce the fit procedure and we show the results. In Section~\ref{sec::DM}, we discuss the DM constraints for both Galactic and extragalactic DM components. Finally, we conclude in Section \ref{sec::concl}. Additionally, we present a phenomenological approach to the interpretation of the UGRB anisotropy energy spectrum in Appendix~\ref{sec::pheno}, while in Appendix~\ref{sec::prev} we relate our results to the findings of previous measurements.

%===================================================================
\section{Modeling blazar populations}
\label{sec::blz}
%===================================================================

In \citep{Manconi:2019ynl} it has been pointed out that a single blazar model is sufficient to describe both the anisotropy level \cp\ and the 4FGL catalog data, at the price of allowing for a relatively broad distribution of the spectral index. Such an approach can be seen as an effective description, where different sub-populations (with narrower spectral index distributions) are combined in a single model \citep{Ajello:2015mfa}. We reproduce the finding of \citep{Manconi:2019ynl}, although in a more general way and by using a phenomenological model, in Appendix \ref{sec::pheno}. However, we note that the blazar model in \citep{Manconi:2019ynl} was only compared to the catalog data in bins of flux and redshift, but not in bins of spectral index which constrains the SED. In contrast, here we intend to use the full catalog information.

In this section, we will therefore consider a physical description of the two populations of blazars that are more numerous in the 4FGL catalog, namely BLLs and FSRQs. We aim to assess their ability to explain the APS measurement. In other words, we test the possibility that FSRQs, with properties compatible with their cataloged sample, are the population that accounts for the low-energy anisotropy found in \citep{Ackermann:2018wlo}, while BLLs are at the origin of the high-energy anisotropy. Below we will show that two populations are preferred when the information of 4FGL catalog are included as a prior.

\subsection{The gamma-ray luminosity function}
%-------------------------------------------------------------------

We summarize here the parametrizations of the gamma-ray luminosity function (GLF) and spectral energy distribution (SED) of BLLs and FSRQs adopted in our analysis. For more details, see~\citep{2012ApJ...751..108A} and \citep{Ajello:2013lka}. The GLF $\Phi(L_{\gamma}, z, \Gamma) = \diff ^3N/\diff L_\gamma \diff V \diff \Gamma$, defined as the number of sources per unit of luminosity $L_{\gamma}$,  co-moving volume $V$ at redshift $z$ and photon spectral index $\Gamma$, is typically decomposed in terms of its expression at $z=0$ and a redshift-evolution function:
\begin{equation}
  \Phi(L_\gamma,z,\Gamma)= \Phi(L_\gamma,0,\Gamma) \times e(L_{\gamma}, z),
  \label{eq:GLF}
\end{equation}
where $L_\gamma$ is the rest-frame luminosity in the energy range $(0.1-100)$ GeV, i.e. $L_\gamma=\int^{100~{\rm GeV}}_{0.1~{\rm GeV}} \diff E_r\,\mathcal{L}(E_r)$ with:
\begin{equation}
  \mathcal{L}(E_r) = \frac{4 \pi d^2_L(z)}{(1+z)} E\,\frac{\diff N}{\diff E}\, ,
  \label{eq:lum}
\end{equation}
$E$ being the observed energy, related to the rest-frame energy $E_r$ as $E_r=(1+z)\,E$. The co-moving volume element in a flat homogeneous Universe is given by $\diff^2V/\diff\Omega \diff z=c\,\chi^2(z)/H(z)$, where $\chi$ is the co-moving distance (related to the luminosity distance $d_L$ by $\chi=d_L/(1+z)$), and $H$ is the Hubble parameter. We use $\Lambda$CDM cosmology with parameters from the final full-mission Planck measurements of the CMB anisotropies~\citep{Planck:2018vyg}. 

At redshift $z=0$, the parametrization of the GLF is: 
\begin{eqnarray}
	\label{eq:glf}
	\Phi(L_\gamma,0,\Gamma) &=& \frac{A}{\ln(10) L_\gamma} \left[ \left( \frac{L_\gamma}{L_0} \right)^{\gamma_1} + 
	\left( \frac{L_\gamma}{L_0} \right)^{\gamma_2} \right]^{-1} \label{eq:GLF_0} \\
	&& \qquad \times
 	\exp \left[ - \frac{(\Gamma-\mu(L_\gamma))^2}{2\sigma^2} \right] \;,\nonumber
\end{eqnarray}
where $A$ is a normalization factor, the indices $\gamma_1$ and $\gamma_2$ govern the evolution of the GLF with the luminosity $L_{\gamma}$ and the Gaussian term takes into account the distribution of the photon indices $\Gamma$ around their mean $\mu(L_{\gamma})$, with a dispersion $\sigma$. It turns out that the GLF of BLLs has a relatively broad distribution in terms of luminosity. For this reason, we allow the mean spectral index to slightly evolve with luminosity from a value $\mu^\ast$: 
\begin{equation}
  \mu(L_\gamma) = \mu^\ast + \beta  
  \left[\log \left(\frac{L_\gamma}{\mathrm{erg~s^{-1}}}\right) - 46\right]\, .
  \label{eqn:mu}
\end{equation}
On the other hand, FSRQs have a narrower distribution, so that the inclusion of this effect would not affect the fit and we can fix $\mu(L_\gamma) = \mu^\ast$.

We adopt a luminosity-dependent density evolution (LDDE):
\begin{eqnarray}
  e(L_{\gamma}, z) =&&  \left[\left(\frac{1+z}{1+z_c(L_\gamma)}\right)^{-p_1} \right. \\
  && \qquad \left. + \left(\frac{1+z}{1+z_c(L_\gamma)}\right)^{-p_2}\right]^{-1}\nonumber
\end{eqnarray}
with 
$z_c(L_\gamma) = z_c^* \times (L_\gamma/10^{48} \mathrm{erg~s^{-1}})^\alpha$, 
$p_1(L_\gamma) = p_1^\ast +\tau \times (\log(L_\gamma) -46)$, and 
$p_2(L_\gamma) = p_2^\ast +\delta \times (\log(L_\gamma) -46)$.
We set $\delta$ to 0.64 for both populations \citep{Ajello:2015mfa}, while $\tau$ is fixed to 3.16 for FSRQs and to 4.62 for BLLs \citep{Ajello:2013lka}.

The SED is modeled as a power law:
\begin{equation}
  \frac{\diff N}{\diff E} \sim \left(\frac{E}{E_0}\right)^{-\Gamma} \;.
\end{equation}
For definiteness, the spetral index $\Gamma$ will be taken to be in the range $(1,3.5)$ (see also \cite{Manconi:2019ynl}). Given the SED,  the photon flux $S(E_{\rm min},E_{\rm max})$ in a given energy interval is obtained by: 
\begin{equation}
  S(E_{\rm min},E_{\rm max}) = \int_{E_{\rm min}}^{E_{\rm max}} \frac{\diff N}{\diff E}   {e^{-\tau(E\,,z)}} \; \diff E,
  \label{eq:flux}
\end{equation}
where $\tau(E,z)$ describes the attenuation by the extragalactic background light (EBL) \citep{Finke2010}. Unless explicitly stated otherwise, the flux in the following computations of the \dnds\ and the associated figures always refer to the energy bin from 1 GeV to 100 GeV. 
The free parameters of the model are summarized in Table~\ref{tab:GLFModel}, together with their best-fit values obtained as outlined below.

With the physical models of the GLF and SED at hand, we can compute the differential number of blazars per integrated flux and solid angle as:
\begin{equation}
  \frac{\diff N}{\diff S} = \int_{0.01}^{5.0} \diff z \int_1^{3.5} \diff \Gamma \, 
  \Phi[ L_\gamma(S,z,\Gamma),z,\Gamma] \, \frac{\diff V}{\diff z} \, 
  \frac{\diff L_\gamma}{\diff S},
  \label{eq:dnds}
\end{equation}
and the size of the gamma-ray intensity fluctuations between energy bins $i$ and $j$ can be cast in the following form (assuming the Poisson-noise term is the dominant contribution):
\begin{eqnarray}
  C_{\rm P}^{ij} &=& \int_{0.01}^{5.0} \diff z\frac{\diff V}{\diff z} \int_1^{3.5} \diff\Gamma 
  \int_{L_{\rm min}}^{L_{\rm max}} \diff L_\gamma\, \Phi(L_\gamma,z,\Gamma) \label{eq::cp} \\
   &\times& S_i(L_\gamma,z,\Gamma) \, S_j(L_\gamma,z,\Gamma) \left[ 1-\Omega(S(L_\gamma,z,\Gamma),\Gamma) \right]\;. \nonumber
\end{eqnarray}
The term $\Omega(S, \Gamma)$ accounts for the \Fermi-LAT sensitivity to detect a source, and it is modeled through a step-function becoming equal to one at the flux threshold sensitivity $S_{\rm thr}$ as described in \citep{Manconi:2019ynl} (Section III.B.1). It depends on $\Gamma$ and it includes a nuisance parameter $k_{C_P}$ which accounts for the uncertainty in its description. We checked that a smooth, more realistic sensitivity function only has a negligible effect on the \cp.
The bounds in the $L_{\gamma}$ integration 
are $L_{\rm min}=7\times10^{43}$~erg/s 
and $L_{\rm max}=1\times10^{52}$~erg/s, for BLL, taken from \citep{Ajello:2013lka}, 
and $L_{\rm min}=1\times10^{44}$~erg/s 
and $L_{\rm max}=1\times10^{52}$~erg/s, for FSRQ, taken from \citep{2012ApJ...751..108A}.

The \cp's of BLL and FSRQ are additive, i.e., $C_{\rm P}=C_{\rm P}^{\rm BLL}+C_{\rm P}^{\rm FSRQ}$. We neglect the (multipole-dependent) clustering term (discussed below in the case of DM), since we checked that, in the multipole range of interest, is a few orders of magnitude smaller than the \cp\ term.

\subsection{The source count distribution}
%-------------------------------------------------------------------

%=====================
%    \                                           |
%      \                                         |
%        \                                       |
\begin{figure}[b!]
  %\centering
    \includegraphics[width=1.0\linewidth,trim={1.4cm 0.5cm 2cm 1cm},clip]{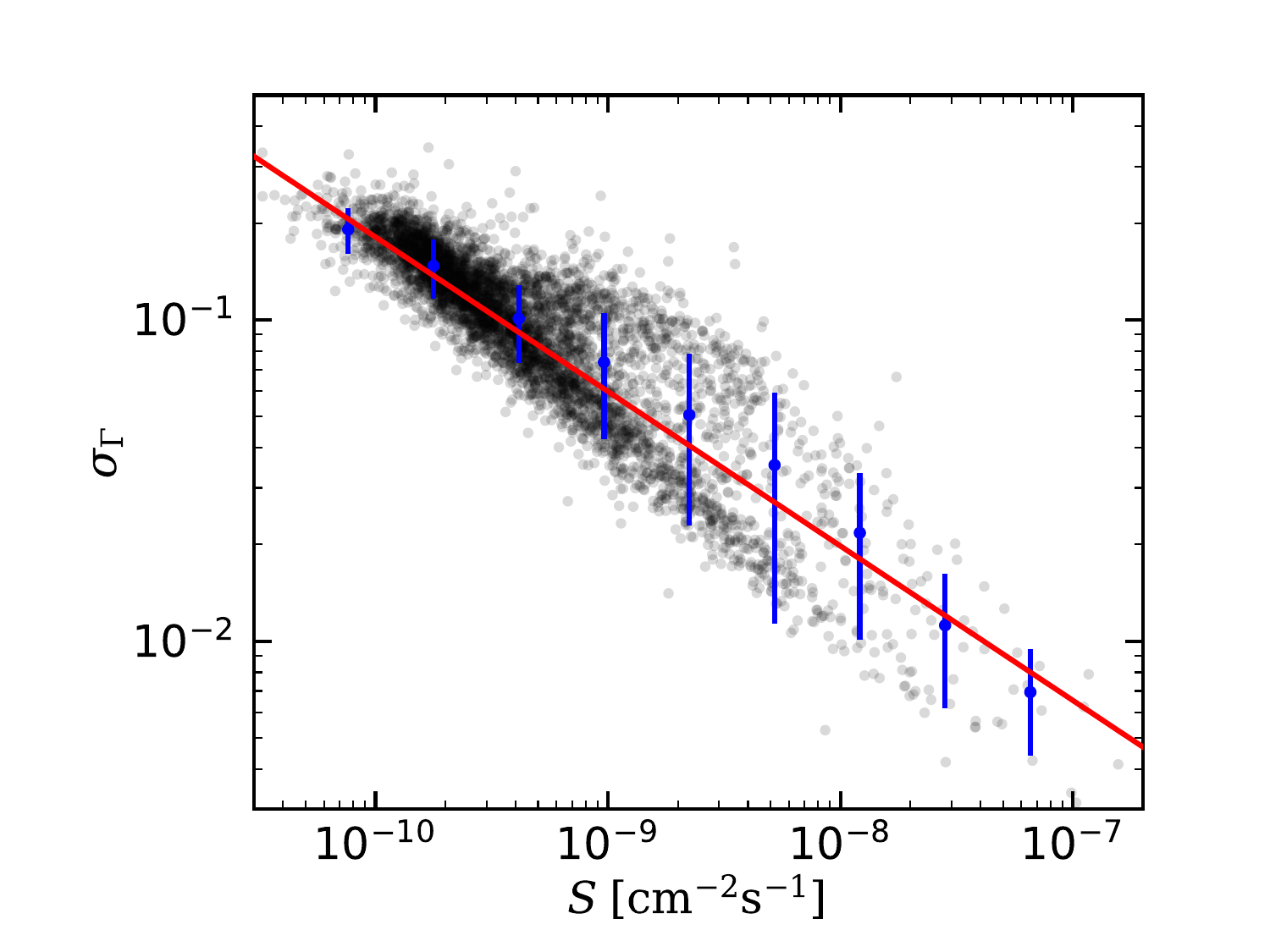}
    \caption{  
               Uncertainty on the determination of the photon spectral index as function of flux. 
            }
    \label{fig::sigma_Gamma}
\end{figure}
%                                      \         |
%                                        \       |
%                                          \     |
%=====================

The source count distribution, $\diff N / \diff S$, is defined as the number of sources per flux and solid angle and is a function of the flux $S$. In principle, there could also be a directional dependence, but blazars are observed up to relatively large distances, such that their distribution can be taken isotropic. On the other hand, populations of blazars are known to evolve with time, \ie\ they depend on redshift. Furthermore, blazars do not have a unique SED: for this reason, we adopt a distribution for the photon spectral index $\Gamma$, which in equation~\eqref{eq:glf} is assumed to be Gaussian.

The source count distribution (see equation~\eqref{eq:dnds}) depends on flux, photon spectral index, and redshift. The latter has been estimated for some of the \Fermi-LAT resolved sources, for which the association to a source from another catalog was possible. We, therefore, build a three-dimensional grid for the average source count distribution in the flux bin $i$, the redshift bin $j$, and the photon spectral index bin $k$:

\begin{eqnarray}    
    \label{eq:dnds_model_bin}
    && \left\langle \frac{\diff N}{\diff S} \right\rangle_{ijk}  = \frac{1}{S_{i,\mathrm{max}}-S_{i,\mathrm{min}}}
     \;\\ \nonumber
    && \qquad \times \!\!
    \int\limits_{     S_{i,\mathrm{min}}}^{     S_{i,\mathrm{max}}} \!\!\!\diff S
    \int\limits_{     z_{j,\mathrm{min}}}^{     z_{j,\mathrm{max}}} \!\!\!\diff z 
    \int\limits_{\Gamma_{k,\mathrm{min}}}^{\Gamma_{k,\mathrm{max}}} \!\!\!\diff \Gamma \, 
    \Phi(L_\gamma,z,\Gamma) \, \frac{\diff V}{\diff z} \, 
    \frac{\diff L_\gamma}{\diff S} \, .
\end{eqnarray}

This is, however, the \emph{true} theoretical prediction of the \dnds. In practice, we have to consider uncertainties of the source parameters as detected by the \Fermi-LAT. This especially matters for those parameters for which the \dnds \, shows a strong dependence. In our case, the first-order behavior of the \dnds \, is a power law as function of $S$, a smoothly broken power law as function of $z$, and a Gaussian as function of $\Gamma$. While the power-law behavior is very smooth, the impact of resolution can be important in the Gaussian tails of $\Gamma$. We use a data-driven method to obtain the uncertainty on the determination of the spectral index. In Figure~\ref{fig::sigma_Gamma} we show the uncertainty of the spectral index, $\sigma_\Gamma$, as stated in the 4FGL catalog against the flux value $S$. As expected, the average uncertainty increases at smaller flux values. To obtain the average uncertainty as function of $S$ we determine the mean and RMS of $\sigma_\Gamma$ in 10 flux bins logarithmically spaced between $5\times10^{-11}\;\mathrm{cm^{-2}s^{-1}}$ and $1\times10^{-7}\;\mathrm{cm^{-2}s^{-1}}$ (blue data points in Figure~\ref{fig::sigma_Gamma}) and then fit a power law (red line) to those data points:
\begin{eqnarray}    
    \label{eq:sigma_Gamma}
    \sigma_\Gamma(S) = A \, S^{-\gamma}.
\end{eqnarray}
The best-fit values are $A=2.82\times 10^{-6}$ and $\gamma = 0.48$.
Assuming that to a first approximation the resolution behaves as a Gaussian, we can convolute the \dnds\ of equation~\eqref{eq:dnds_model_bin} with a Gaussian with the width $\sigma_\Gamma(S)$. The inclusion of the experimental resolution is a key ingredient to obtain a reasonable fit for small fluxes and extreme $\Gamma$ bins. However, this would lead to a quadruple integral for the \dnds\, which is computationally not feasible. So, instead of integrating over the flux bin, we simply evaluate the \dnds\ at the (geometric) mean, $S_i$, of the flux bin which gives a good approximation because of the smooth behavior of the \dnds. So finally for the comparison with the 4FGL catalog, the \dnds\ of our model is given by:
\begin{eqnarray}    
    \label{eq:dnds_model_bin_inc_fermi_resolution}
    && \!\!\!\left\langle \frac{\diff N}{\diff S} \right\rangle_{ijk}  =
    \int\limits_{     z_{j,\mathrm{min}}}^{     z_{j,\mathrm{max}}} \!\!\!\diff z 
    \int\limits_{\Gamma^{\rm (obs)}_{k,\mathrm{min}}}^{\Gamma^{\rm (obs)}_{k,\mathrm{max}}} \!\!\!\diff \Gamma^{\rm (obs)} \, 
    \int \diff\Gamma \, \frac{\diff V}{\diff z} \, 
    \frac{\diff L_\gamma}{\diff S} \,  \\ \nonumber
    && \qquad \qquad \times
    \frac{\exp\left(-\frac{1}{2}\left(\frac{\Gamma-\Gamma^{\rm (obs)}}{\sigma_\Gamma (S_i)}\right)^2\right)}{\sqrt{2\pi}\sigma_\Gamma(S_i)} \,
    \Phi\left(L_\gamma(S_i,z),z,\Gamma\right) \, .
\end{eqnarray}

\subsection{Fit procedure}
%-------------------------------------------------------------------

Our data sets are the resolved sources collected in the \Fermi-LAT 4FGL catalog \citep{Fermi-LAT:2019yla}, from which we construct the source count \dnds, and the anisotropy of the gamma-ray sky due to unresolved sources, encoded in the $C_{\rm P}^{ij}$ and measured in \citep{Ackermann:2018wlo}. We perform two sets of fits: one on the 4FGL catalog alone and one which combines the 4FGL catalog with the \cp\ measurement. The fit strategy follows the procedure introduced in \citep{Manconi:2019ynl}. 

In this paper, we model the luminosity function of BLLs and FSRQs separately. We assume that both blazar populations can be described by the functional form of equation~\ref{eq:glf}, but with different parameter values. 
The log-likelihood of our fit to the 4FGL catalog data and the \cp\ data is given by the sum of the two individual $\chi^2$s:
\begin{eqnarray}
	\label{eqn::logL}
    -2&&\log\mathcal{L}(\boldsymbol \theta_{\rm BLL},\boldsymbol \theta_{\rm FSRQ},\boldsymbol \theta_{\rm n}) = \\ \nonumber
         && \chi^2_{\mathrm{4FGL}}(\boldsymbol \theta_{\rm BLL},\boldsymbol \theta_{\rm FSRQ}) 
         + \chi^2_{C_{\rm P}}(\boldsymbol \theta_{\rm BLL},\boldsymbol \theta_{\rm FSRQ},\boldsymbol \theta_{\rm n}).
\end{eqnarray}
Here $\boldsymbol \theta_{\rm BLL}$ and $\boldsymbol \theta_{\rm FSRQ}$ denote the parameters of the BLL and FSRQ models, while $\boldsymbol \theta_{\rm n}$ marks nuisance parameters for the \cp. In the following we also use a short notation: $\boldsymbol \theta_{\rm blz} = \lbrace \boldsymbol \theta_{\rm BLL}, \boldsymbol \theta_{\rm FSRQ}, \boldsymbol \theta_{\rm n} \rbrace$. We note that in practice there is only a single nuisance parameter, $k_{C_\mr{P}}$.

\subsubsection{Fit on the 4FGL catalog data}
%-------------------------------------------------------------------

We extract the source count distribution of four different source classes from the 4FGL catalog and then test our model against these distributions. In more detail, we use the following four source classes:%
\begin{itemize}
    \item 
    BLL: the sum of {\sl identified} and {\sl associated BL} Lacs (sources labeled \texttt{BLL} or \texttt{bll} in the 4FGL catalog),
    \item 
    FSRQ: the sum of {\sl identified} and {\sl associated} FSRQs (sources labeled \texttt{FSRQ} or \texttt{fsrq} in the 4FGL catalog),
    \item 
    BLZ: the sum of {\sl identified} and {\sl associated} blazars  (sources labeled \texttt{BLL}, \texttt{BCU}, \texttt{FSRQ}, \texttt{bll}, \texttt{bcu}, or \texttt{fsrq} in the 4FGL catalog), and
    \item
    ALL: {\sl all sources} in the catalog.
\end{itemize}
We expect our model to respect the following constraints:
\begin{itemize}
    \item 
    The sum of the source count distribution of our models for BLLs and FSRQs
    should not overshoot the observed total source count distribution of all sources (ALL).
    In this sense, the source count distribution ALL provides an upper bound for our models.
    This contribution to the $\chi^2$ is labeled $\chi^2_{\rm ALL}$.
    \item 
    The sum of the source count distribution of our models for BLLs and FSRQs
    should be at least as large as the sum of identified and associated blazars (BLZ).
    However, our models are allowed to lie above the observed source count distribution 
    because of unassociated sources. In this sense, BLZ provides a lower bound for our models 
    (labeled $\chi^2_{\rm BLZ}$).
    \item Our BLL model has to explain at least the observed source count distribution 
    BLL and thus also provides a lower bound (labeled $\chi^2_{\rm BLL}$).
    \item In analogy to BLL, also the FSRQ model receives a lower bound from the 
    observed source count distribution (labeled $\chi^2_{\rm FSRQ}$). 
\end{itemize}
Each of these constraints would give rise to a contribution of the 4FGL $\chi^2$. However, a naive sum of the $\chi^2$ from the three lower bounds would lead to a double counting since BLL and FSRQ sources appear also in the BLZ class. To avoid this, we consider only the most constraining lower bound between the BLZ case and the combination of BLL and FSRQ, i.e.,
\begin{eqnarray}
	\label{eqn::chi2_4FGL}
	\chi^2_{\mathrm{4FGL}} = \chi^2_{\mathrm{ALL}} 
	+ \max\left( \chi^2_{\mathrm{BLZ}}, \chi^2_{\mathrm{BLL}} + \chi^2_{\mathrm{FSRQ}} \right).
\end{eqnarray}

\medskip

The upper bound from ALL is implemented as:
\begin{eqnarray}
	\label{eqn::chi2_4FGL_1_test}
	\chi^2_{\mathrm{ALL}} =
	  \sum\limits_i 
	    \left[\max \left( \frac{ \left\langle \frac{\diff N}{\diff S}\right\rangle_{{\rm BLZ},i} - \left(       \frac{\diff N}{\diff S}      \right)_{\mathrm{ALL},i} 
	                       }
	          {   \sigma_{\mathrm{ALL},i}   }  ,  0 \right) \right]^2
\end{eqnarray}
Here the \dnds\ in angle brackets denotes the model prediction from equation~\eqref{eq:dnds_model_bin_inc_fermi_resolution} and the \dnds\ in round brackets is the source count distribution extracted from the 4FGL catalog. The model includes the sum of BLL and FSRQ, namely $\langle \diff N/\diff S \rangle_{{\rm BLZ},i}=\langle \diff N/\diff S \rangle_{{\rm BLL},i}+\langle \diff N/\diff S \rangle_{{\rm FSRQ},i}$. For this contribution, we integrate over all redshifts and all values of the photon spectral index. The remaining index $i$ denotes the flux bin, as summarized in Table~\ref{tab:4FGL_binning}. If the flux of the bin is below the detection threshold $S_\mathrm{thr}$, the bin is excluded from the sum.

%=====================
%    \                                           |
%      \                                         |
%        \                                       |
\begin{deluxetable*}{cccccc}
\tablenum{1}
\caption{
	Binning of the 4FGL fit.
	\label{tab:4FGL_binning}
	}
\tablewidth{700px}
\tablehead{
Source class & variable & min & max & number of bins & scaling  
}
\startdata
ALL & $S\;\mathrm{[cm^{-2}s^{-1}]}$ & $10^{-10}$ & $10^{-7}$ & 10 & log             \\
    & $\Gamma$                      & 1.0        & 3.5       &  1 & linear          \\
BLZ & $S\;\mathrm{[cm^{-2}s^{-1}]}$ & $10^{-10}$ & $10^{-7}$ & 10 & log             \\
    & $\Gamma$                      & 1.0        & 3.5       &  1 & linear          \\
    & $z$                           & 0.0        & 4.0       &  6 & log in $(1+z)$  \\
BLL & $S\;\mathrm{[cm^{-2}s^{-1}]}$ & $10^{-10}$ & $10^{-7}$ & 10 & log             \\
    & $\Gamma$                      & 1.6        & 2.4       &  5 & linear          \\
    & $z$                           & 0.0        & 4.0       &  6 & log in $(1+z)$  \\
FSRQ& $S\;\mathrm{[cm^{-2}s^{-1}]}$ & $10^{-10}$ & $10^{-7}$ & 10 & log             \\
    & $\Gamma$                      & 2.1        & 2.9       &  5 & linear          \\
    & $z$                           & 0.0        & 4.0       &  6 & log in $(1+z)$  \\
\enddata
\end{deluxetable*}
%                                      \         |
%                                        \       |
%                                          \     |
%=====================

%=====================
%    \                                           |
%      \                                         |
%        \                                       |
\begin{deluxetable*}{ccclcc}
\tablenum{2}
\caption{
	Fit results of the GLFs.
	\label{tab:GLFModel}
	}
\tablewidth{700px}
\tablehead{
$                                                                      $& \multicolumn2c{4FGL fit}                          &\ \ \ & \multicolumn2c{4FGL$+$\cp\ fit}                        \\
\cline{2-3}
\cline{5-6}           
$                                                                      $& FSRQ                       & BLL                        && FSRQ                       & BLL                                    
}
\startdata
$\log_{10}\left({A}_{\mathrm{}}\;\mathrm{[Mpc^{-3}]}\right)            $&$ -9.35_{-  0.37}^{+  0.62}$&$ -9.80_{-  1.11}^{+  0.40}$&&$ -9.65_{-  0.47}^{+  0.61}$&$ -9.01_{-  1.11}^{+  1.21}$\\
$\log_{10}\left({L^*}_{\mathrm{}}\;\mathrm{[erg/s]}\right)             $&$ 48.36_{-  0.66}^{+  0.31}$&$ 47.85_{-  0.52}^{+  0.79}$&&$ 48.53_{-  0.60}^{+  0.42}$&$ 47.26_{-  1.09}^{+  0.75}$\\
${\gamma_1}_{\mathrm{}}                                                $&$  0.57_{-  0.09}^{+  0.15}$&$  1.03_{-  0.07}^{+  0.12}$&&$  0.72_{-  0.09}^{+  0.13}$&$  0.92_{-  0.07}^{+  0.18}$\\
${\gamma_2}_{\mathrm{}}                                                $&$  1.93_{-  0.43}^{+  0.14}$&$  1.95_{-  0.45}^{+  0.16}$&&$  1.97_{-  0.42}^{+  0.21}$&$  1.88_{-  0.38}^{+  0.12}$\\
${z_c^*}_{\mathrm{}}                                                     $&$  0.93_{-  0.27}^{+  0.20}$&$  1.05_{-  0.53}^{+  0.16}$&&$  0.87_{-  0.24}^{+  0.14}$&$  1.06_{-  0.56}^{+  0.16}$\\
${p_1^*}_{\mathrm{}}                                                   $&$  5.86_{-  5.02}^{+  2.15}$&$  7.48_{-  4.51}^{+  3.97}$&&$  8.37_{-  3.35}^{+  3.78}$&$  4.01_{-  3.54}^{+  0.77}$\\
${p_2^*}_{\mathrm{}}                                                   $&$ -0.88_{-  0.14}^{+  0.77}$&$ -1.97_{-  0.43}^{+  1.83}$&&$ -0.77_{-  0.11}^{+  0.67}$&$ -0.93_{-  0.19}^{+  0.83}$\\
${\alpha}_{\mathrm{}}                                                  $&$  0.20_{-  0.16}^{+  0.07}$&$  0.28_{-  0.13}^{+  0.14}$&&$  0.11_{-  0.11}^{+  0.02}$&$  0.31_{-  0.07}^{+  0.17}$\\
${\mu^*}_{\mathrm{}}                                                     $&$  2.50_{-  0.04}^{+  0.03}$&$  2.03_{-  0.04}^{+  0.04}$&&$  2.49_{-  0.04}^{+  0.03}$&$  2.05_{-  0.04}^{+  0.03}$\\
${\sigma}_{\mathrm{}}                                                  $&$  0.19_{-  0.04}^{+  0.02}$&$  0.22_{-  0.05}^{+  0.03}$&&$  0.20_{-  0.04}^{+  0.02}$&$  0.19_{-  0.03}^{+  0.02}$\\
${\beta}_{\mathrm{}}                                                   $&                            &$  0.06_{-  0.05}^{+  0.02}$&&                            &$  0.03_{-  0.03}^{+  0.01}$\\
$k_{C_\mathrm{P}}                                                      $&                            &                            && \multicolumn2c{$  1.09_{-  0.12}^{+  0.16}$}            \\
\hline
$\chi^2_\mathrm{ALL      }                                             $& \multicolumn2c{ 1.7}                                    && \multicolumn2c{ 2.9}                                    \\
$\chi^2_\mathrm{BLZ      }                                             $& \multicolumn2c{ 3.0}                                    && \multicolumn2c{ 2.7}                                    \\
$\chi^2_\mathrm{BLL      }                                             $& \multicolumn2c{11.4}                                    && \multicolumn2c{12.7}                                    \\
$\chi^2_\mathrm{FSRQ     }                                             $& \multicolumn2c{ 7.4}                                    && \multicolumn2c{ 8.0}                                    \\
$\chi^2_\mathrm{4FGL     }                                             $& \multicolumn2c{20.5}                                    && \multicolumn2c{26.3}                                    \\
$\chi^2_{C_\mathrm{P}}                                                 $& \multicolumn2c{    }                                    && \multicolumn2c{81.0}                                    \\
$\chi^2                                                                $& \multicolumn2c{20.5}                                    && \multicolumn2c{107.3}                                   \\
\enddata
\end{deluxetable*}

For the lower bounds, related to identified or associated blazars, we would like to consider also the redshift information, which is not directly provide in the 4FGL catalog. So, we extract the information from the 4LAC catalog \citep{Fermi-LAT:2019pir}, which contains spectroscopic redshift measurements. However, the redshift information of the 4LAC catalog is incomplete, \emph{i.e.} not all sources have a redshift measurement. Since we consider the identified or associated blazars only as a lower bound this does not represent a problem but it might not be the most constraining option. Hence, we consider again two cases. 
In the first case, we include redshift information and compare our model in bins of a two-dimensional grid in redshift and flux through:
\begin{eqnarray}
	\label{eqn::chi2_4FGL_2_1}
	\!\!\!\!\!\!\!\chi^2_{{\rm BLZ},Sz} =
	  \sum\limits_{i,j} 
	    \left[\min \left( \frac{   \left\langle \frac{\diff N}{\diff S}\right\rangle_{{\rm BLZ},ij} 
	                             - \left(       \frac{\diff N}{\diff S}      \right)_{{\rm BLZ},ij} }
	        {  \sigma_{\mathrm{BLZ},ij}   } ,  0 \right) \right]^2
\end{eqnarray}
In the second case, we disregard the redshift information by integrating over the redshift. We define:
\begin{eqnarray}
\label{eqn::chi2_4FGL_2_2}
	\chi^2_{{\rm BLZ},S} =
	  \sum\limits_i 
	    \left[\min \left( \frac{   \left\langle \frac{\diff N}{\diff S}\right\rangle_{{\rm BLZ},i} 
	                             - \left(       \frac{\diff N}{\diff S}      \right)_{{\rm BLZ},i} }
	        {  \sigma_{\mathrm{BLZ},i}   } ,  0 \right) \right]^2
\end{eqnarray}
Depending on the model parameter point, either equation~\eqref{eqn::chi2_4FGL_2_1} or equation~\eqref{eqn::chi2_4FGL_2_2} provides the stronger constraint. Similar to the discussion above we cannot use the sum of both $\chi^2$s because of double counting, and again we choose the most constraining one:
\begin{eqnarray}
	\label{eqn::chi2_BLZ}
	\chi^2_{\mathrm{BLZ}} = 
	\max \left(\chi^2_{\mathrm{BLZ},Sz},\chi^2_{\mathrm{BLZ},S}\right) .
\end{eqnarray}

As described in equation~\eqref{eqn::chi2_4FGL}, we select the lower bounds by comparing $\chi^2_{\mathrm{BLZ}}$ with the ones from the analysis of the individual source classes BLL and FSRQ. In this latter case, we can use the full information of the catalogs and compare models with data on a three-dimensional grid in flux, redshift, and photon spectral index. Again, since the redshift information is incomplete, we define the $\chi^2$ as the maximum of the two cases:
\begin{eqnarray}
	\label{eqn::chi2_ML}
	\chi^2_{\mathrm{M}} = 
	\max \left(\chi^2_{\mathrm{M},Sz\Gamma},\chi^2_{\mathrm{M},S\Gamma}\right) ,
\end{eqnarray}
where $M$ stands for either BLL or FSRQ. The individual $\chi^2$
in the two cases are defined by
\begin{eqnarray}
\label{eqn::chi2_4FGL_3_1}
	\chi^2_{{\rm M},Sz\Gamma} =
	  \sum\limits_{i,j,k}  
	    \left[\min \left( \frac{   \left\langle \frac{\diff N}{\diff S}\right\rangle_{{\rm M},ijk}
	                             - \left(       \frac{\diff N}{\diff S}      \right)_{{\rm M},ijk} }
	        {  \sigma_{{\rm M},ijk}   } ,  0 \right) \right]^2
\end{eqnarray}
when redshift information is included, and
\begin{eqnarray}
\label{eqn::chi2_4FGL_3_2}
	\chi^2_{{\rm M},S\Gamma} =
	  \sum\limits_{i,k}  
	    \left[\min \left( \frac{   \left\langle \frac{\diff N}{\diff S}\right\rangle_{{\rm M},ik} 
	                             - \left(       \frac{\diff N}{\diff S}      \right)_{{\rm M},ik} }
	        {  \sigma_{{\rm M},ik}   } ,  0 \right) \right]^2
\end{eqnarray}
in the case with flux and spectral index bins only.

To avoid a bias from the Galactic plane we exclude small latitudes with $|b|< 30$ deg from our analysis.

%=====================
%    \                                           |
%      \                                         |
%        \                                       |
\begin{figure*}[t]
    \setlength{\unitlength}{1\textwidth}
    \begin{picture}(1,0.7)
      \put(0.00 ,-0.03){\includegraphics[width=0.5\textwidth]{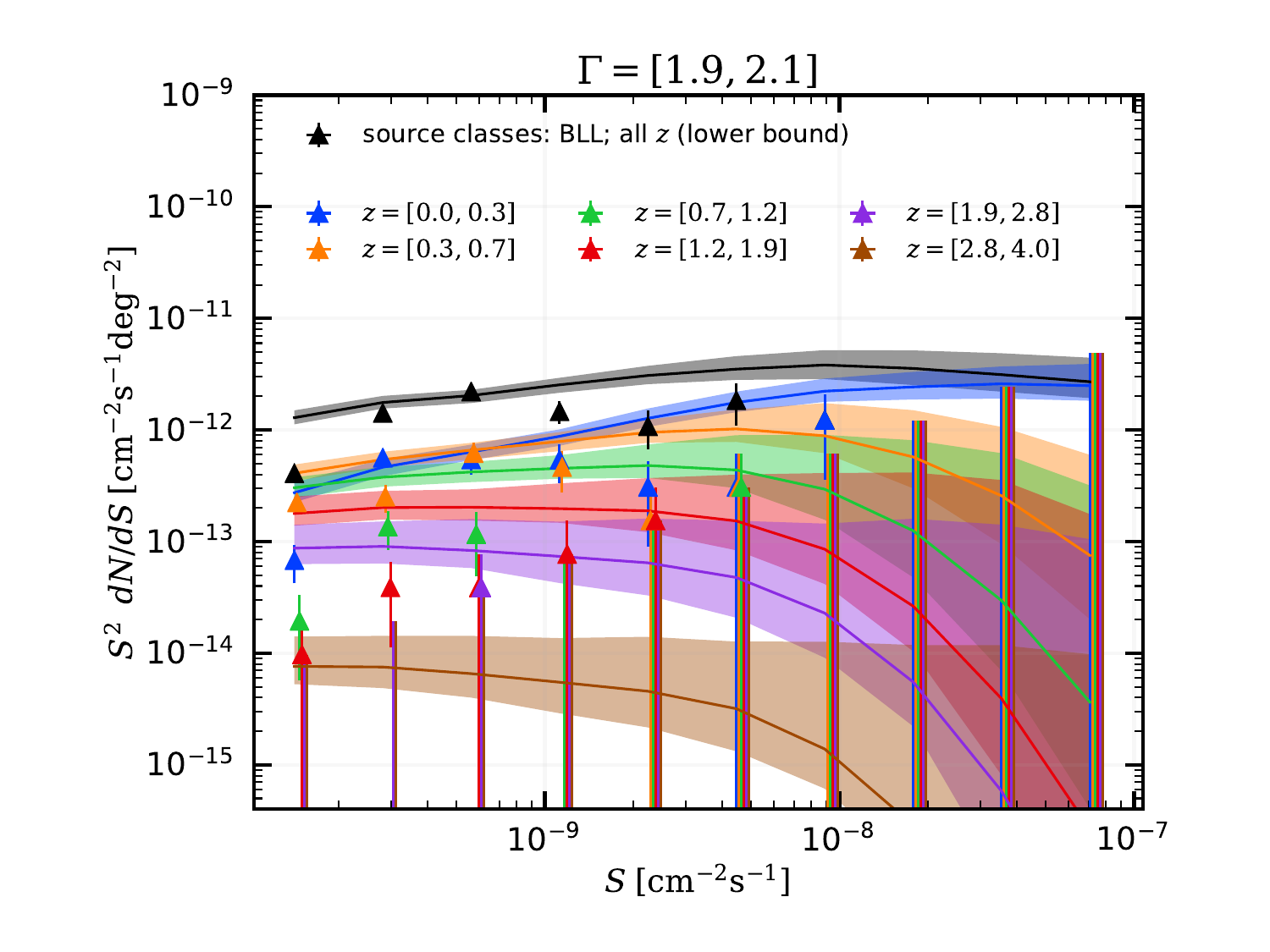}}
      \put(0.50 ,-0.03){\includegraphics[width=0.5\textwidth]{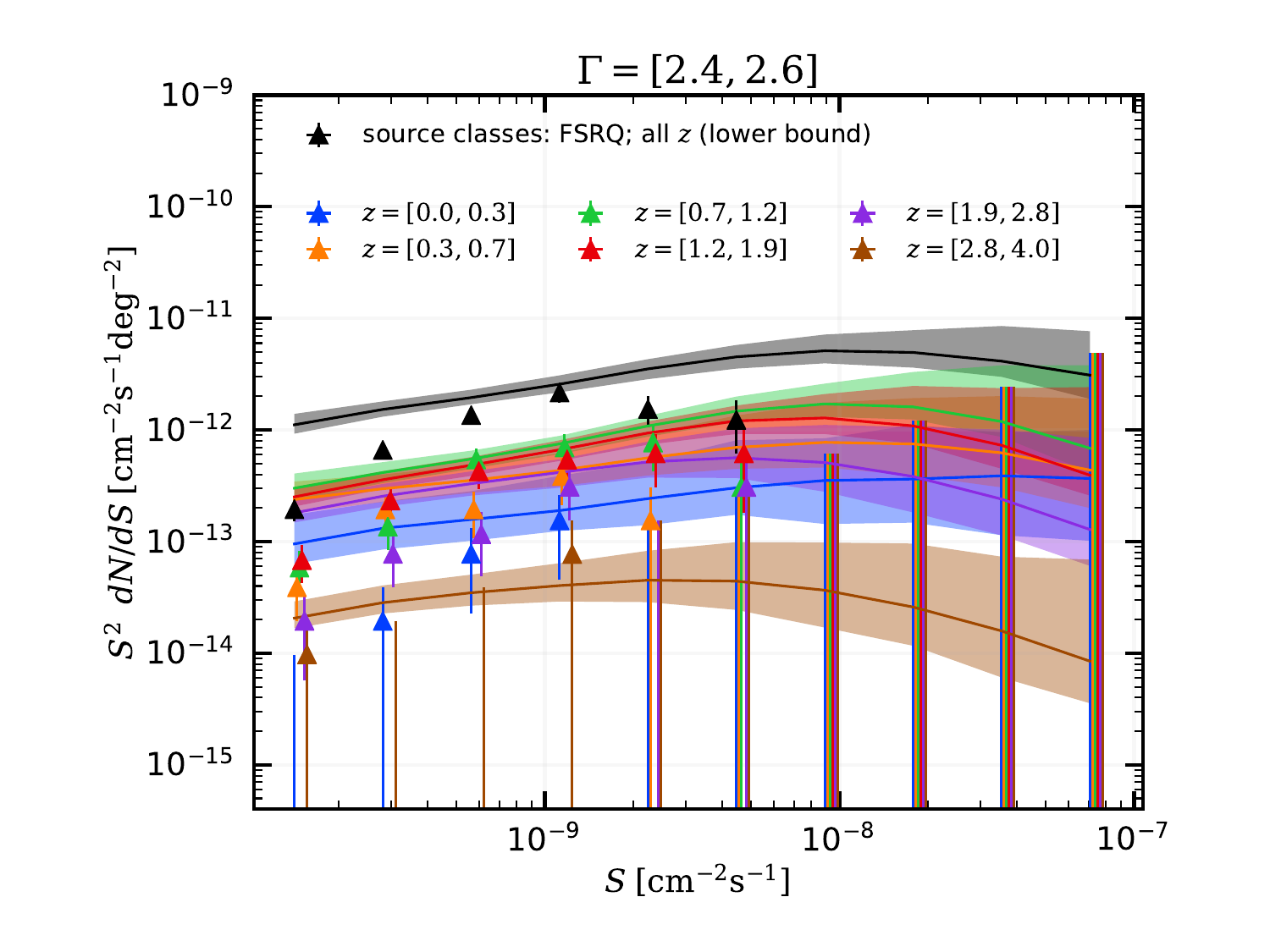}}
      \put(0.00 , 0.33){\includegraphics[width=0.5\textwidth]{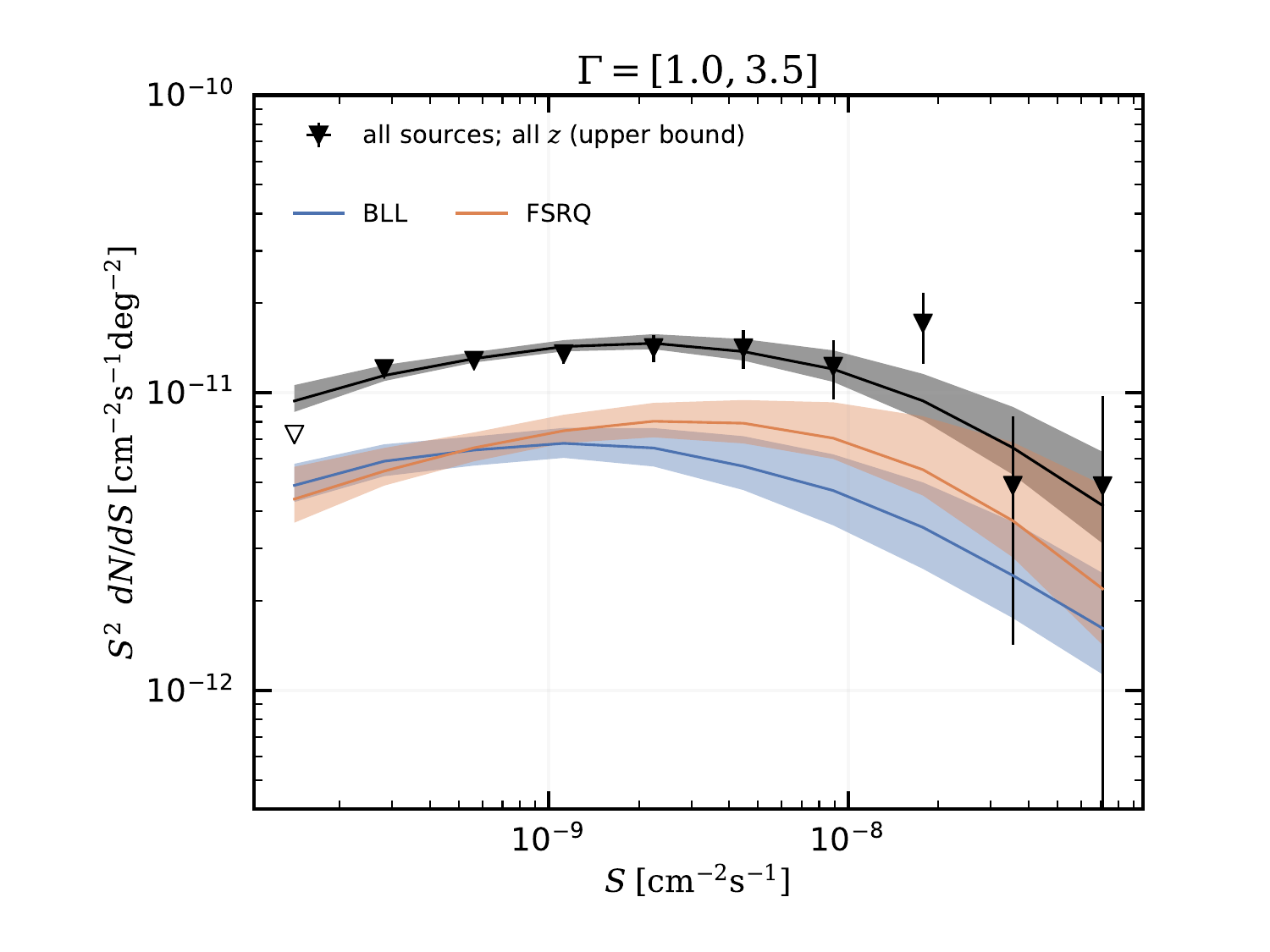}}
      \put(0.50 , 0.33){\includegraphics[width=0.5\textwidth]{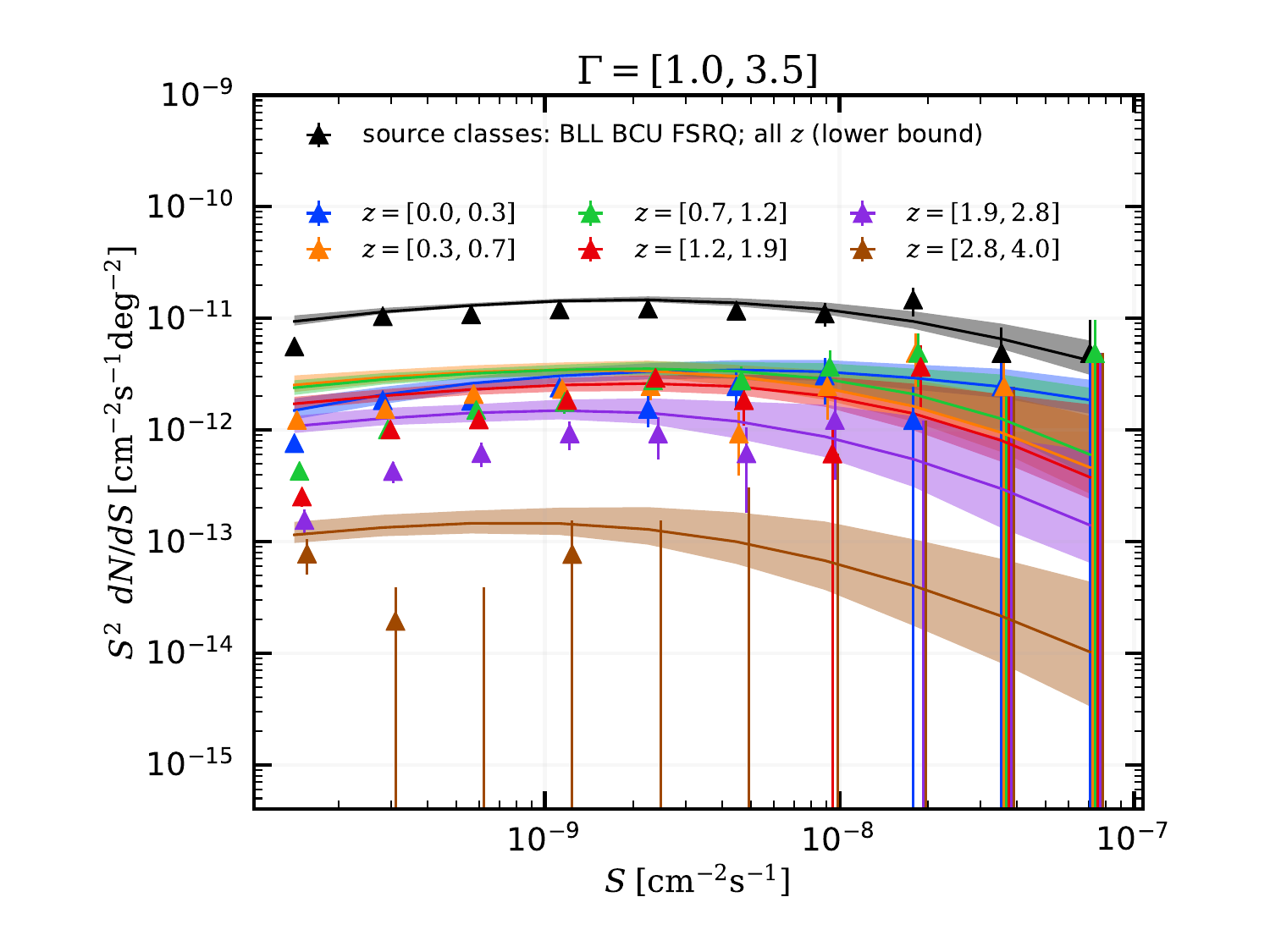}}
      %\put(0.00 , 0.33){\includegraphics[width=0.5\textwidth]{test.pdf}}
      \put(0.00 , 0.33){\includegraphics[width=0.5\textwidth]{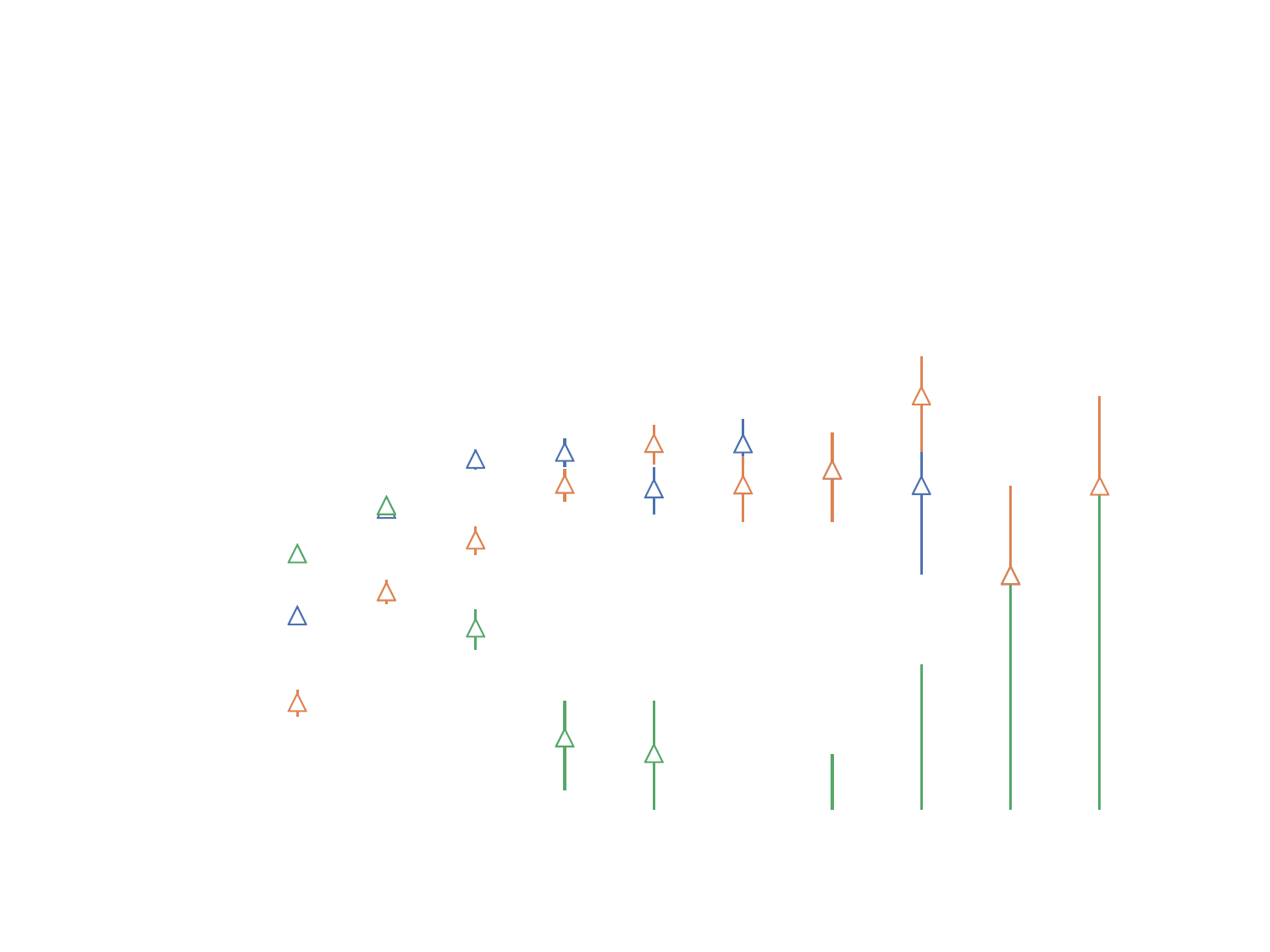}}
    \end{picture}
    \caption{  
         	 Source count distribution of the 4FGL sources in bins of flux ($S$), photon spectral index ($\Gamma$), and redshift ($z$). The GLF is fitted to 4FGL$+C_\mathrm{P}$. 
	         The band shows the $1\sigma$ Bayesian uncertainty.
	         The black open (filled white) data point in the upper left panel is below the flux threshold and thus not included in the fit (see text for further details). In our statistical analysis, the data of the upper-left panel are taken as upper bounds, while for all the other panels they are lower bounds.
	         The colored open data points show the identified BLL (blue), FSRQ (orange), and BCU (green). Those points are not included in the fit and only shown for comparison.
		     \label{fig::fitCp4FGL_best_fit_dNdS}  
            }
\end{figure*}
%                                      \         |
%                                        \       |
%                                          \     |
%=====================

\subsubsection{Fit on the $C_\mathrm{P}$ data}
%-------------------------------------------------------------------

The fit of the APS is performed on the auto ($i=j$) and cross ($i\neq j$) correlation measurements, where $i$ and $j$ denote the energy bins. The $\chi^2_{C_{\rm P}}$ is defined as: 
\begin{eqnarray}
	\label{eqn::chi2_ASP}
	\chi^2_{C_{\rm P}} = \sum\limits_{i\leq j} 
	\frac{  \left[   \left(   C_{\rm P}^{ij}    \right)_{\mathrm{meas}} 
	               - \left(   C_{\rm P}^{ij}    \right)_{\mathrm{th}}     \right]^2   }
	     {   \sigma_{C_{\rm P}^{ij}}^2   } . 
\end{eqnarray}
The subscript \emph{meas} refers the measured \cp\ obtained in \citep{Ackermann:2018wlo}, while the subscript \emph{th} denotes the theoretical estimation of \cp\ calculated as in equation~\eqref{eq::cp}. Finally, $\sigma_{C_{\rm P}^{ij}}^2$ are the uncertainties of the measured \cp, again taken from \citep{Ackermann:2018wlo}.
\bigskip

\subsection{Analysis strategy}
\label{sec:BlazarModelStrategy}
%-------------------------------------------------------------------

As anticipated above, in this work, we perform two fits. The first one utilizes only the 4FGL catalog while the second one additionally uses the \cp\ measurement. The basic idea is as follows: From the first fit, we obtain constraints on the GLF and SED models of the two blazar populations in the flux regime of resolved point sources. From there we can extrapolate to the unresolved flux regime and calculate the \cp. As will become clear in the next Section~\ref{sec:BlazarModelRes}, this extrapolation agrees well with the actual \cp\ measurement. So, we can go one step further and also perform the second fit that combines the resolved point sources (4FGL) with the \cp. This combined fit provides GLF and SED models which are consistent with gamma-ray observations also below the below the flux threshold of the 4FGL catalog. When we derive DM constraints in Section~\ref{sec::DM} the combined fit serves as the baseline.

The large parameter space investigated in this work is sampled using \textsc{MultiNest} \citep{2009MNRAS.398.1601F}. We use a configuration with 800 live points, an enlargement factor of \texttt{efr}~$=0.7$, and a stopping parameter of \texttt{tol}~$=0.1$. In the following, we present the results in the Bayesian statistical framework. 

\subsection{Results on the GLF and SED of BLLs and FSRQs}
\label{sec:BlazarModelRes}
%-------------------------------------------------------------------

As a result of our fits, we obtain the GLFs and SEDs of FSRQs and BLLs. The 4FGL categorization of the blazars into the two classes is incomplete which leaves some degeneracy. We allow the fit to attribute the uncharacterized blazars either to BLLs or FSRQs. This is only possible because we fit both source classes at the same time. We note that this treatment leads to correlations of BLL parameters with FSRQ parameters and vice versa. These correlations are important to assess the uncertainty of the full blazar model correctly, as for example in  Section~\ref{sec::DM} where blazars pose the background for our DM search. 

The results are summarized in Table~\ref{tab:GLFModel}, where we state the mean values and the 1$\sigma$ uncertainty derived from the marginalized posterior for each parameter. Results are provided for two setups: In the first setup, we only fit to the resolved point sources of the 4FGL catalog, while in the second setup we fit both the resolved sources and the APS data. The obtained parameter values of the two setups are compatible within their uncertainties. The GLF parameters of the FSRQ and BLL model are very well compatible with the parameters values of \cite{2012ApJ...751..108A} and \cite{Ajello:2013lka}, respectively. This means that the GLF of sources belonging to the fainter regime probed in this work closely follows from the one in the brighter end. We note that we use a slightly different definition of the LDDE than \cite{2012ApJ...751..108A} which leads to slightly different parameter values for $p_1^\ast$, $p_2^\ast$, and $z_c^\ast$.

%=====================
%    \                                           |
%      \                                         |
%        \                                       |
\begin{figure*}[t!]
    \includegraphics[width=.5\linewidth]{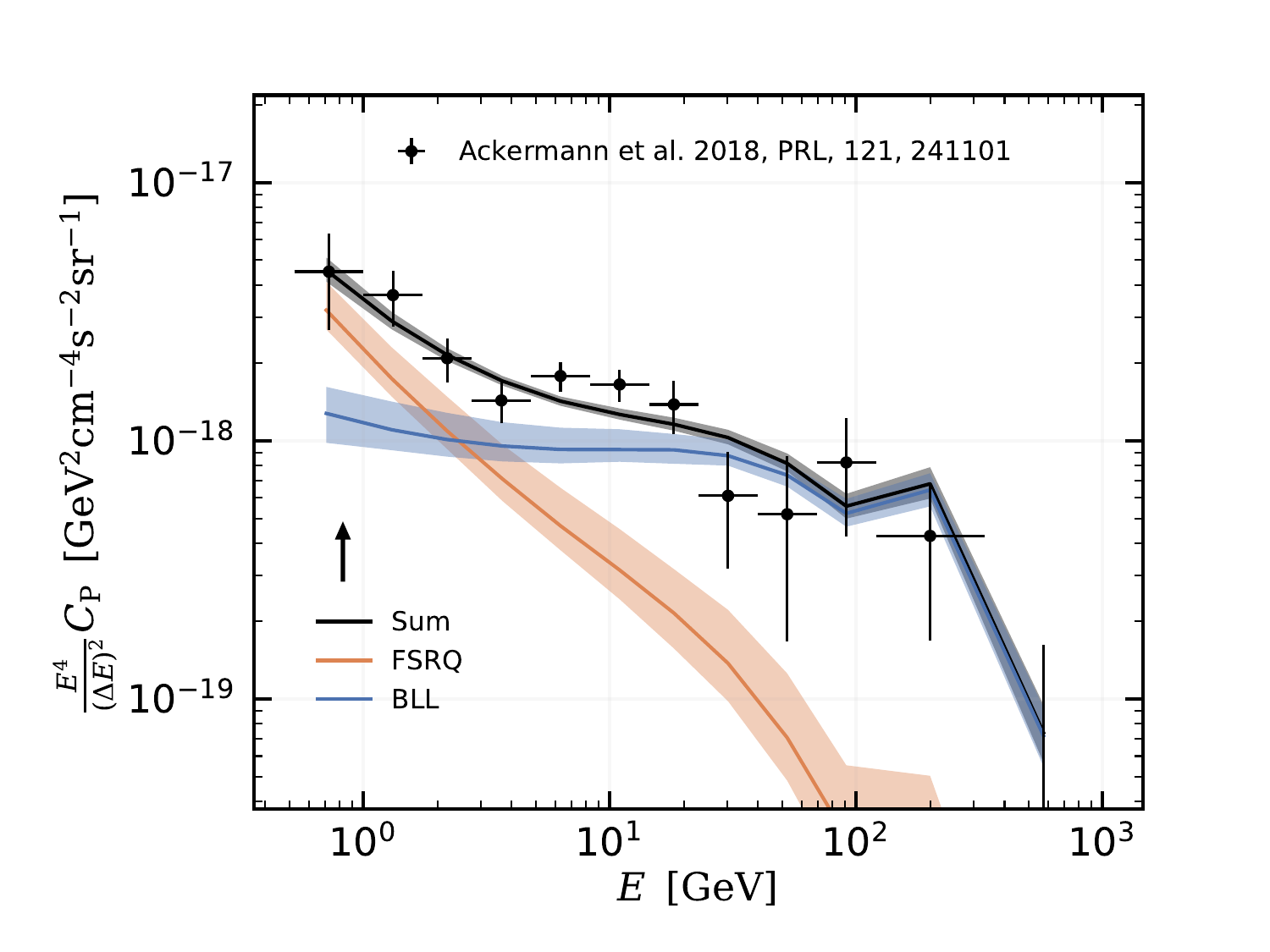}\includegraphics[width=.5\linewidth]{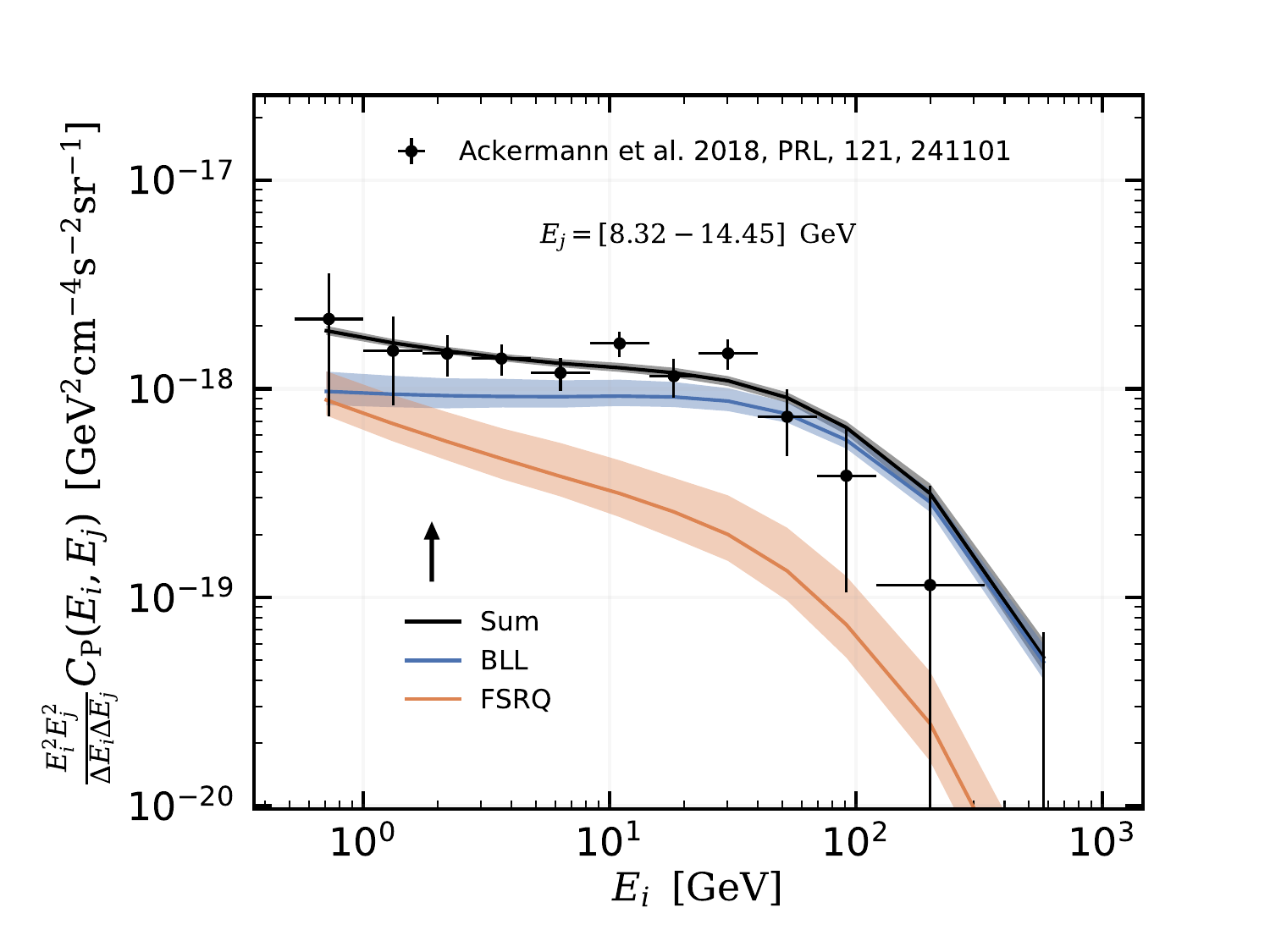}
    \caption{  
         	 Angular correlation of the 4FGL$+C_\mathrm{P}$ fit. The shaded bands mark the the $1\sigma$ Bayesian uncertainty.
		     \label{fig::fitCp4FGL__Cp}  
            }
\end{figure*}
%                                      \         |
%                                        \       |
%                                          \     |
%=====================

Figure \ref{fig::fitCp4FGL_best_fit_dNdS} shows the best-fit and uncertainties of \dnds\ from the combined fit of 4FGL+\cp \, in comparison to the \dnds\ extracted from the 4FGL catalog. The four different panels correspond to the different contributions to $\chi^2_\mr{4FGL}$. In more detail, the data points in the upper left panel show the \dnds\ of all sources from the 4FGL catalog (at  $|b|>30$~deg). The sum of our models for BLLs and FSRQs is in agreement with those data. Because of the statistical technique we adopted (see above), it is expected to stay at the level or below those data points. The open white data point is below the flux threshold, so it is excluded from the analysis. The upper right panel shows the data points of the \dnds\ for all identified or associated blazars. The different colors show the \dnds\ in different redshift bins, while the black points are summed over all redshifts. Our model for the \dnds\ of BLLs plus FSRQs lies as expected at the level or above the data points. In both upper panels, the source count distribution is integrated over all photon spectral indices, from 1.0 to 3.5. Furthermore, the upper panels only show constraints from the 4FGL catalog on the sum of the BLL and FSRQ models. The two lower panels, instead, look at the individual models for FSRQs and BLLs. Furthermore, they focus on the \dnds\ for specific bins of the photon spectral index, corresponding to the peak of the distribution for each class. The lower left panel compares the \dnds\ BLLs in the $\Gamma$ bin from 1.9 to 2.1. Again the different colors correspond to different redshift bins and the black points contain the sum over all redshifts. Finally, the lower right panel is as the left panel but for FSRQs and a $\Gamma$ bin from 2.4 to 2.6. All in all, we see that our model matches the constraints from the 4FGL catalog very well. We show these plots only for the 4FGL+\cp\ fit but we note that they look very similar for the 4FGL-only fit.

%=====================
%    \                                           |
%      \                                         |
%        \                                       |
\begin{figure}[b!]
    \includegraphics[width=1.0\linewidth,trim={1cm 0.9cm 2cm 1cm},clip]{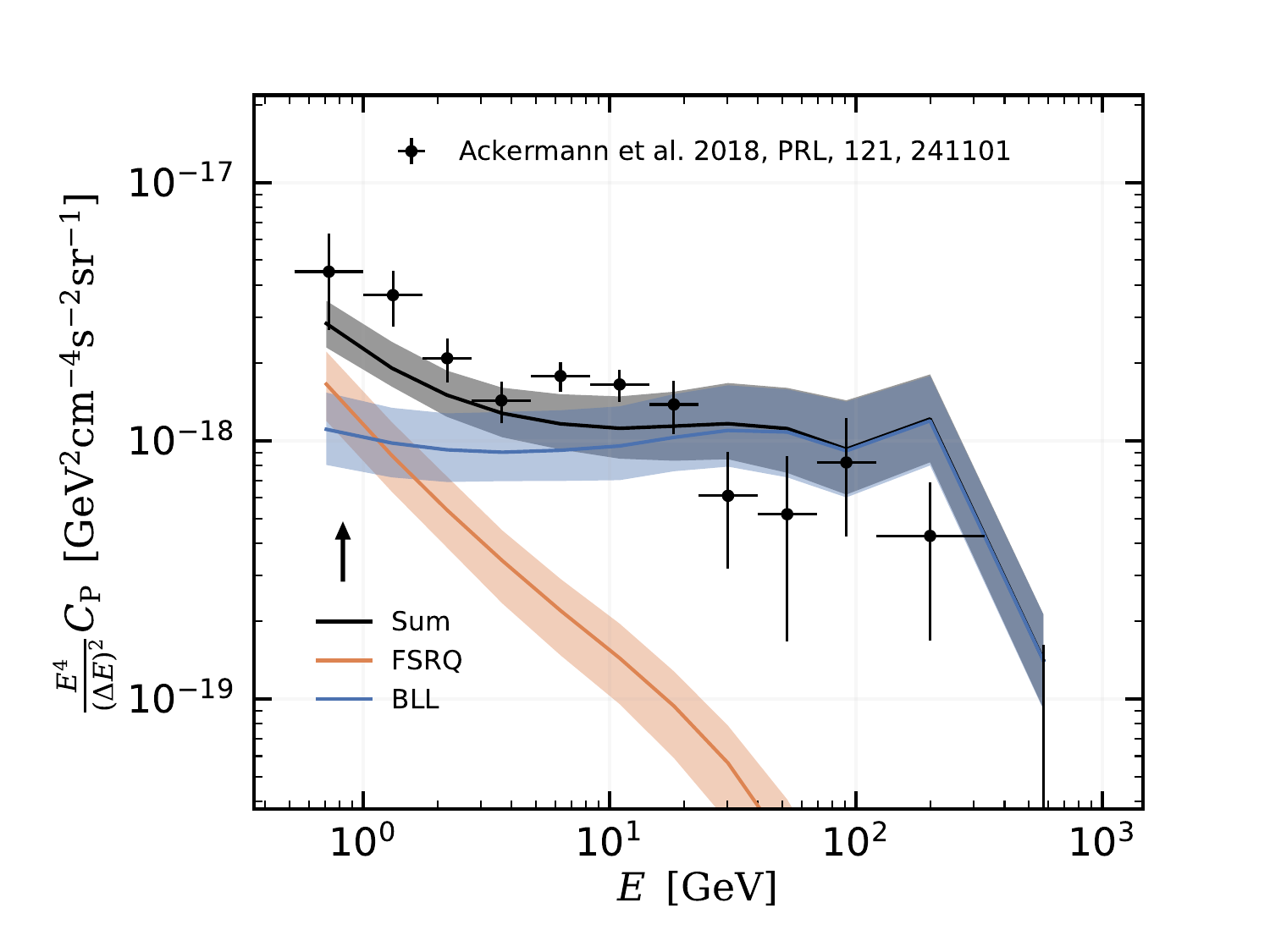}
    \caption{  
         	 The GLF is fitted to 4FGL sources and then extrapolated to the \cp. The band shows the $1\sigma$ Bayesian uncertainty. 
		     \label{fig::fit4FGL__Cp_diagonal}  
            }
\end{figure}
%                                      \         |
%                                        \       |
%                                          \     |
%=====================
%=====================
%    \                                           |
%      \                                         |
%        \                                       |
\begin{figure}[b!]
    \includegraphics[width=1.0\linewidth,trim={1cm 0.1cm 1.5cm 1cm},clip]{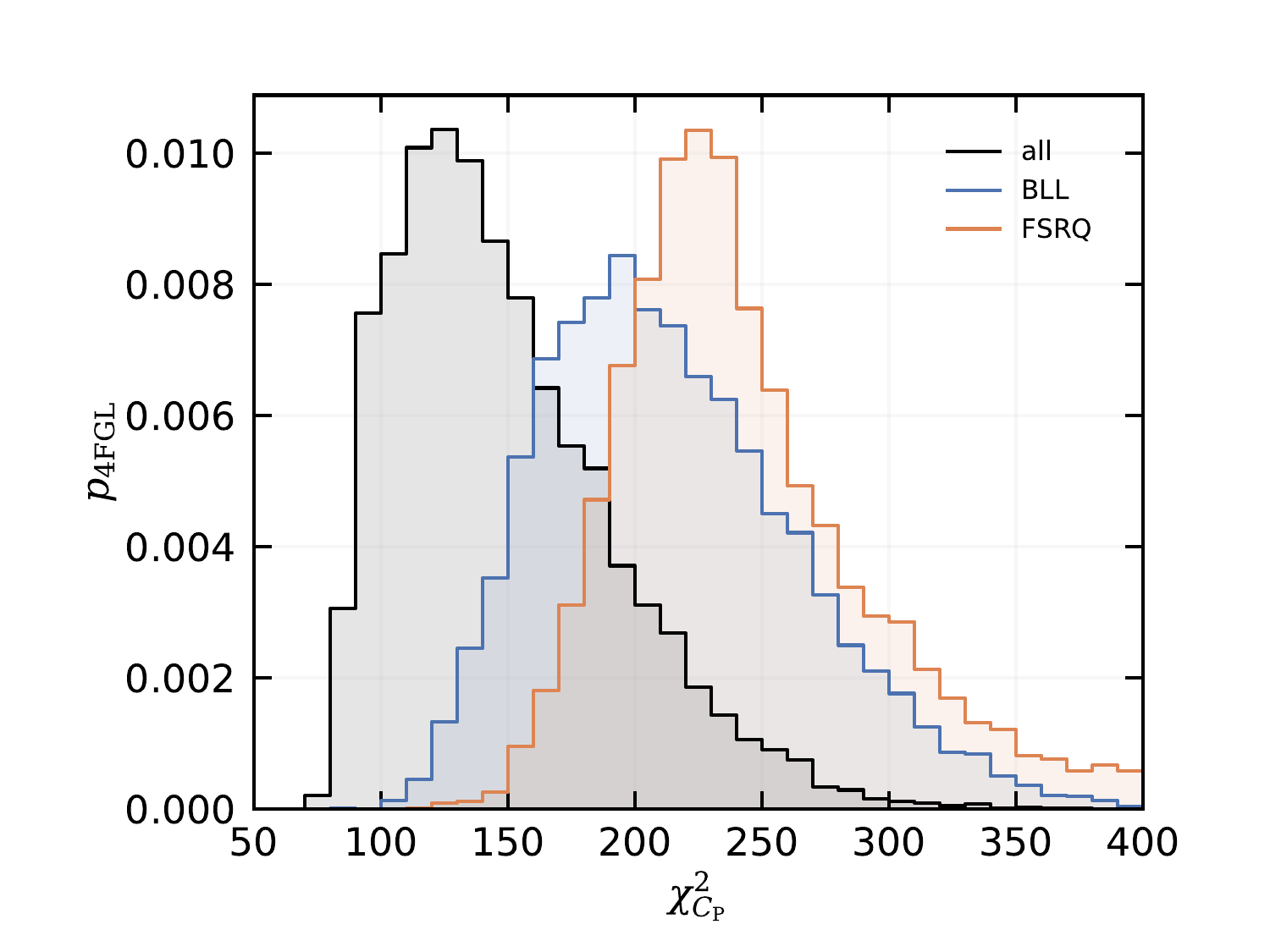}
    \caption{  
         	 Posterior distribution of the 4FGL-only fit for the $\chi^2_{C_\mr{P}}$. We show the distributions for three cases: considering the sum of BLLs and FSRQs (all), only BLLs (BLL), and only FSRQs (FSRQ). 
		     \label{fig::fit4FGL__Cp_chi2_posterior}  
            }
\end{figure}
%                                      \         |
%                                        \       |
%                                          \     |
%=====================

%=====================
%    \                                           |
%      \                                         |
%        \                                       |
\begin{figure*}[t]
    \includegraphics[width=1.0\linewidth,,trim={5cm 5cm 5cm 5cm},clip]{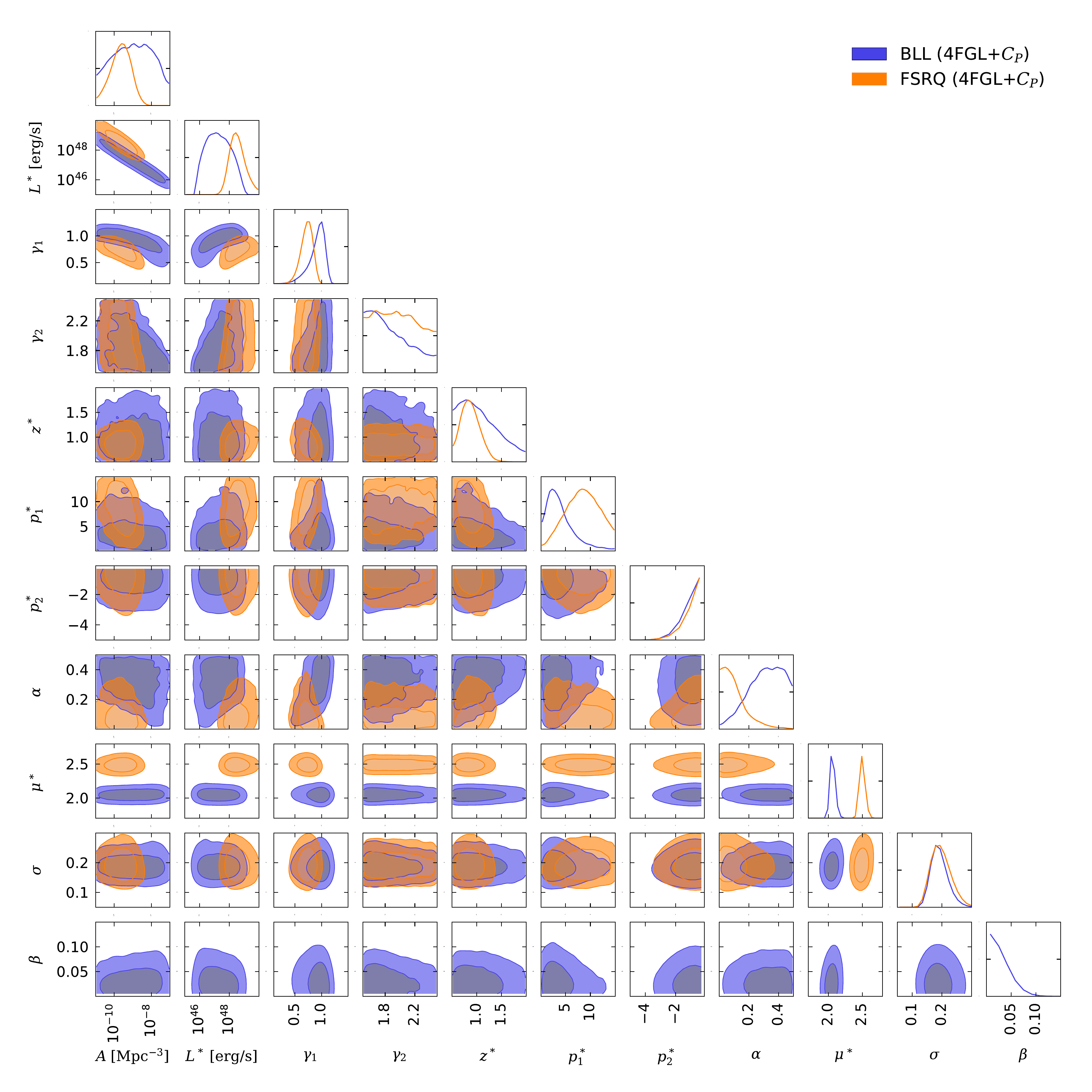}
    \caption{  
         	  Triangle showing the parameters constraints of the GLF of FSRQs (amber) and BLL (blue). Both constraints are derived from a to 4FGL$+C_\mathrm{P}$ data.
         	  The panels on the diagonal show the marginalized posterior distribution of each single parameter, while the panels in the lower half show the 1 and 2$\sigma$ uncertainty contours derived from the tow-dimensional marginalized posterior for each combination of two parameters.
		     \label{fig::triangle_FSRQvsBLL}  
            }
\end{figure*}
%                                      \         |
%                                        \       |
%                                          \     |
%=====================

Figure \ref{fig::fitCp4FGL__Cp} compares the \cp\ measurement with the best fit model and uncertainty from the 4FGL+\cp\ setup. The left panel shows the \cp\ auto-correlation, while the right panel shows an example of cross-correlation, between the (8.3, 14.5)~GeV energy bin and all other energy bins. The sum of BLL and FSRQ model provides a good fit to the data. This was also the case in the analysis of unresolved sources after 2 years of \Fermi-LAT data \cite{Mauro_2014}. However, in the meantime many more point sources were resolved and the \cp\ measurement decreased by a factor of about 10 making the latest data much more sensitive to faint populations. Still we find that blazars explain the entire latest \cp\ measurement. The feature at 200 GeV in our model (and also, less visible, in the data) is related to a change in the analysis adopted for the measurement from \citep{Ackermann:2018wlo}. While for energies below 200 GeV bright point sources are masked from the 4FGL catalog, in the last two bins at high energies, bright point sources were masked using the 3FHL catalog. This leads to a change of the actual $S_\mr{thr}$ which explains the feature. The feature does not appear in the right panel because for the cross-correlation the masks of the two energy bins involved in the measurement were joined (and so when an energy bin below 200 GeV is present, the mask is mostly provided by point sources of the 4FGL catalog). Note also from Table~\ref{tab:GLFModel} that the nuisance parameter $k_{C_\mr{P}}$, introduced to allow for a possible rescaling of the flux threshold sensitivity from the reference model, is within uncertainties compatible with the default value of 1.

It is also interesting to look at the setup of 4FGL-only, and use the extrapolation of the GLF and SED model to predict the \cp. As shown in Figure~\ref{fig::fit4FGL__Cp_diagonal} our prediction agrees very well with the measurement. We note that the \cp\ is dominated by FSRQs at low energies, below $\sim2$~GeV, while BLLs dominate at higher energies. The domination of BLLs at high energies is expected since they have a harder SED than FSRQs. The fact that there is a transition from FSRQs to BLLs in the \cp\ at low energies introduces a softening in the spectral index at low energies. This softening has previously been interpreted as a possible hint for a new source population \citep{Ando:2017alx}. The new \cp\ data \citep{Ackermann:2018wlo} and the detailed treatment presented here, allow us to interpret it in terms of FSRQ. 

Figure \ref{fig::fit4FGL__Cp_diagonal} already hints that both BLLs and FSRQs are required to describe the \cp\ data. We quantify this statement a bit better in the following. Using the results of the 4FGL-only fit, we calculate the posterior distribution of the $\chi^2_{C_{\rm P}}$ assuming (i) the sum of FSRQs and BLLs, labeled all, and (ii) only BLLs. An only FSRQs hypothesis is excluded since it cannot explain the high-energy \cp\ data. In order to consider the systematic uncertainty on the exact flux threshold, we profile over the normalization of the $C_p$. The results in Fig.~\ref{fig::fit4FGL__Cp_chi2_posterior} show that the sum of FSRQ and BLL is preferred. Using the full posterior of the 4FGL-only fit and defining $\mathcal{L}_{C_{\rm p}} = \exp(-\chi^2_{C_{\rm p}}/2)$, we calculate the Bayes factors of the hypotheses (i) and (ii) obtaining $4.0\times10^3$. More details about the calculation of the Bayes factors are given in App.~\ref{sec::bayes_factor_Cp}. In this sense, the two physical populations (BLLs and FSRQs) are clearly preferred over a single population of only BLLs or FSRQs.
We note, however, that the preference for these two population is based on the catalog prior. As discussed in the Appendix \ref{sec::pheno}, the \cp\ data by itself is not sufficient to distinguish between a scenario with one or two populations. Without the catalog prior, a hypothetical and more general GLF and SED can provide a good fit to the \cp\ data.

Finally, Figure~\ref{fig::triangle_FSRQvsBLL} shows a triangle plot with posterior distributions for parameters of the GLF and SED for the BLL and FSRQ models. The posteriors correspond to the 4FGL+\cp\ setup. The diagonal contains the marginalized one-dimensional posteriors for each individual parameter, while the panels in the lower half show the 1 and 2$\sigma$ contours for each combination of two parameters. Since the BLL and FSRQ models have the same functional form we can combine them into the same triangle, using different colors. We observe that the SED parameters $\mu^\ast$ and $\sigma$ are well constrained for both populations. The average photon spectral index of BLLs is $\mu^\ast \sim 2.0$ and a width of $\sigma \sim 0.2$. As expected, FSRQs follow a softer energy spectrum with an average index with $\mu^\ast\sim2.5$ but a similar width. Also, the shape of the GLF at small $L$ is reasonably constrained. The index $\gamma_1$ lies between 0.4 and 1.0 for FSRQs and between 0.5 and 1.2 for BLLs, while the behavior at large $L$ (see $\gamma_2$) is less constrained. We note the degeneracy between $A$ and $L^{\ast}$. This is because, at first order, in both cases, their main impact in the fit is to change the normalization of the GLF. The redshift dependence is only weakly constrained due to degeneracies with other parameters. 

We provide the covariance matrix of our fits in the ancillary files (arXiv version). This covers, to a first approximation, the degeneracies and correlations of the fit parameters. As a final comment, let us note that, clearly, the uncertainties on the parameters of BLLs and FSRQs show some level of mutual correlation in the fit. It is just for the sake of clearness that we do not show the entire triangle plot in Figure~\ref{fig::triangle_FSRQvsBLL}.

\subsection{Blazar contribution to the UGRB}
\label{sec::UGRB}
%-------------------------------------------------------------------

\begin{figure}[t]
    \includegraphics[width=1.0\linewidth,trim={1cm 0.9cm 2cm 1cm},clip]{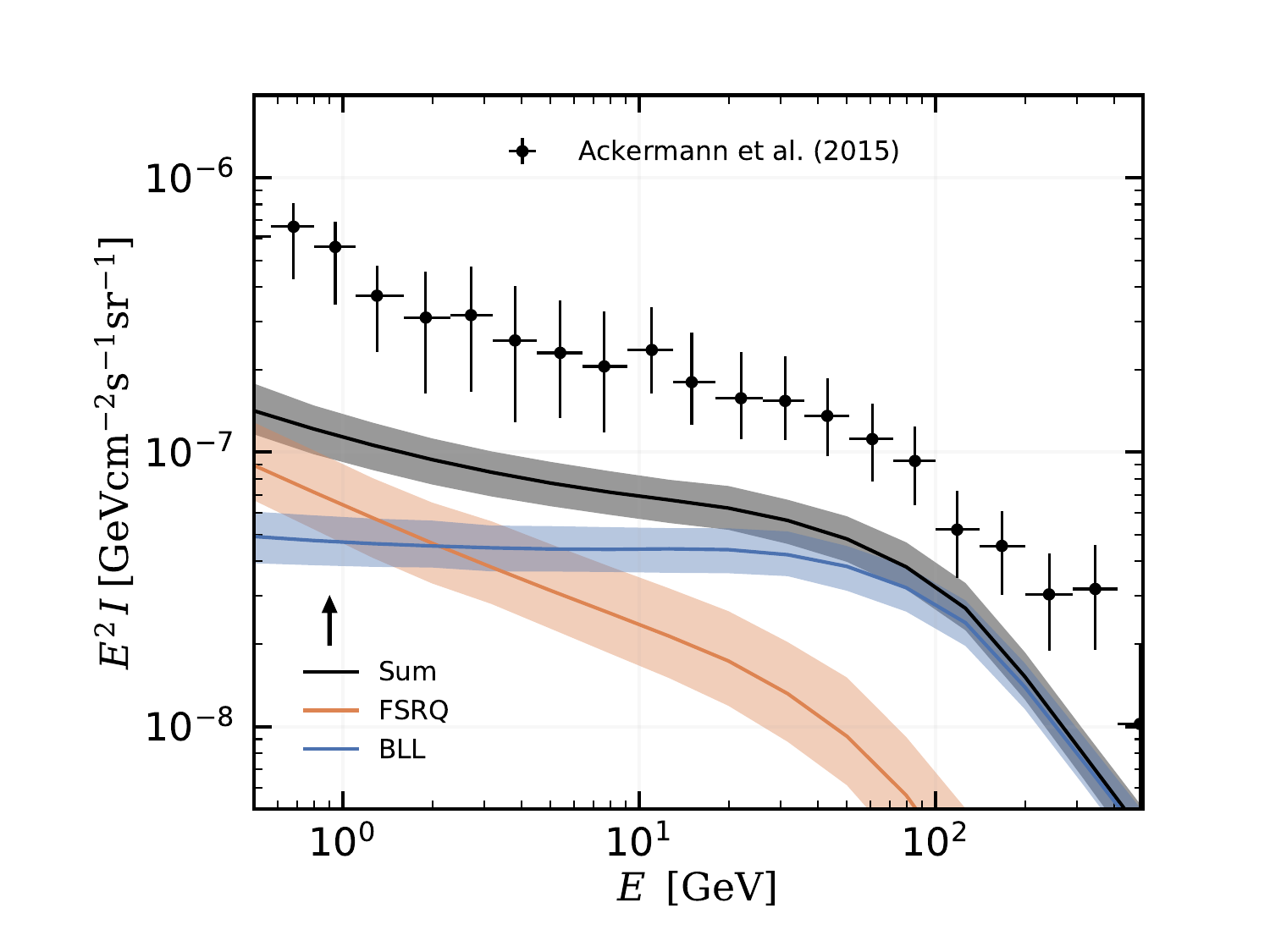}
    \caption{  
         	 Contribution of BLLs and FSRQs to the UGRB intensity. The flux threshold corresponds to the 2FGL catalog, both for the data points and the model prediction. The band shows the $1\sigma$ Bayesian uncertainty. 
		     \label{fig::fit4FGL__UGRB}  
            }
\end{figure}

We see that the measured gamma-ray angular correlations require the presence of FSRQs and BLLs. Populations with a GLF peaked at lower luminosities (like misaligned AGN and star forming galaxies) cannot account for the \cp\ data. Thus, FSRQs and BLLs provide an unavoidable contribution to the UGRB intensity. In Figure~\ref{fig::fit4FGL__UGRB}, we compare the UGRB measurement of the \Fermi-LAT from \cite{Ackermann_2015} with the prediction from our models. The measurement accounted for contribution of point sources from the 2FGL catalog \citep{2012ApJS..199...31N}. To be consistent, we apply a flux threshold corresponding to the 2FGL catalog, taken from \cite{2015ApJ...810...14A}, for the predictions in Figure~\ref{fig::fit4FGL__UGRB}. We conclude that blazars provide a significant contribution to the UGRB, accounting for about 30\% between 10 and 100 GeV. At energies below 1 GeV the contribution decreases to about 20\%.

%=====================
%    \                                           |
%      \                                         |
%        \                                       |
%                                      \         |
%                                        \       |
%                                          \     |
%=====================

%===================================================================
\section{Bounds on WIMP dark matter}
\label{sec::DM}
%===================================================================

The UGRB observed by the \Fermi-LAT could conceal a signal from DM particles. We focus our analysis on annihilating dark matter.
Since annihilation occurs universally in all dark matter structures, we need to consider the gamma-ray emission from both the halo of our Galaxy and from extragalactic structures. We, therefore, have two dark matter contributions to the anisotropy APS, one for the Galactic halo and one from the extragalactic dark matter distribution (the two contributions are not expected to cross-correlate). Moreover, extragalactic structures host the same astrophysical sources (BLL and FSRQ) which we have discussed in the previous sections. This induces a cross-correlation term in the APS between extragalactic dark matter and sources emission. All these terms are properly modeled and taken into account in our analysis, as outlined below. In the following, we will discuss the contribution of DM only to the anisotropy signal \cp, since the contribution of dark matter halos to the source count distribution is significantly suppressed as compared to the one arising from astrophysical sources in the resolved flux regime. A DM contribution to the \dnds\ could in principle emerge on top of astrophysical sources only at very low fluxes.

\subsection{Extragalactic Dark Matter modeling}
%-------------------------------------------------------------------

The APS of the cross-correlation between a source-field $X$ in the energy bin $i$ and a source-field $Y$ in the energy bin $j$ reads \citep{Fornengo2014}
\begin{equation}
  C_{\ell}^{X_iY_j} = 
  \int\!\! \frac{\diff \chi}{\chi^2} W^X_i(\chi)W^Y_j(\chi) 
  P^{X_iY_j}\!\!\left(k=\frac{\ell}{\chi},\chi\right),
  \label{eqn:APS}
\end{equation}
where $W^{{X}}(\chi)$ is the window function of field $X$, $P^{{XY}}(k, \chi)$ the 3-dimensional cross power-spectrum of the fluctuations of the two fields and $\chi$ denotes the comoving distance, related to redshift by $d\chi = (c\,dz)/H(z)$.

The window function for annihilating DM is given by \citep{Ando2006,Fornengo2014}
\begin{eqnarray}
	W_{\rm DM}(z,E) &=& \frac{(\odm \rho_c)^2}{4\pi} \frac{\langle\sv\rangle}{2\mdm^2} \left(1+z\right)^3 \Delta^2(z) \nonumber\\
					&\times& \frac{\de N_{\rm ann}}{\de E}\left[E(1+z) \right] e^{-\tau\left[z,E(1+z)\right]} \ ,
	\label{eq:win_annDM}
\end{eqnarray}
where $\odm$ and $\rho_c$ are the present-day cosmological abundance of DM and the critical density of the Universe, respectively, $\mdm$ is the mass of the DM particle, $\Delta^2(z)$ is the clumping factor, and $\langle\sv\rangle$ denotes the velocity-averaged annihilation cross-section of dark matter particles, assumed here to be the same in all dark matter halos. $\de N_{\rm ann} / \de E$ indicates the number of photons produced per annihilation as a function of energy and sets the gamma-ray energy spectrum and $\tau(E, z)$ denotes the optical depth of gamma-ray photons, which we model as in \citep{2010ApJ...712..238F}.

To determine the DM auto-correlation, we compute the 3D power spectrum with the so-called halo-model approach (see, e.g., the review in \cite{Cooray2002}). For $X=Y=\mathrm{DM}$, we have
\begin{eqnarray}
    P_{{\rm DM},{\rm DM}}^\mr{1h}(k,z)& =& \int_{M_{\rm min}}^{M_{\rm max}} \de M\ \frac{\de n_\mr{h}}{\de M} \,\left(\frac{ \hat u_\kappa(k|M,z)}{\Delta^2(z)}\right)^2 \nonumber\\
    P_{{\rm DM},{\rm DM}}^\mr{2h}(k,z) &=& \left[\int_{M_{\rm min}}^{M_{\rm max}} \de M\,\frac{\de n_\mr{h}}{\de M}\, b_\mr{h} \,\frac{ \hat u_\kappa(k|M,z)}{\Delta^2(z)} \right]^2\nonumber\\
    &\times&  P^{\rm lin}(k,z)\ ,
	\label{eq:PSDM}
\end{eqnarray}
where $\de n_{\rm h}/\de M$ is the halo mass function, $P^{\rm lin}(k,z)$ is the linear matter power spectrum, $b_\mr{h} (M)$ is the linear bias, and $\hat u_{\rm ann}(k|M,z)$ denotes the Fourier transform of the density profile of the DM halos (see, e.g., see the Appendix of \cite{Cuoco2015}). We assume the NFW DM density profile \citep{Navarro:1995iw}. All the ingredients in equations~(\ref{eq:win_annDM}) and (\ref{eq:PSDM}) are modeled as in \citep{DES:2019ucp}. The minimal and maximal halo masses are set at $M_{\rm min} = 10^{-6} M_\odot$ and $M_{\rm min} = 10^{18} M_\odot$.

To characterize the halo profile and the subhalo contribution, we need to specify their mass concentration. The description of the concentration parameter $c(M,z)$ at small masses and for subhalos is still an open issue and provides our largest source of uncertainty. In the following, we consider two models that we name `LOW', where we take the description of $c(M,z)$ from \citep{Correa:2015dva}, and `HIGH', where we follow \citep{Neto:2007vq}. They differ for what concerns the extrapolation of the concentration at low masses and lead to a difference of about one order of magnitude in the final bounds on the annihilation cross section, as shown in Section \ref{subsec::DMcons}.

Clearly, the distribution of extragalactic DM halos (and in turn of the annihilation signal) has some level of correlation with the blazar distribution, therefore inducing a cross-correlation signal between DM and blazars. The blazar window function can be phrased as: $W_{\rm BLA}(z,E) = \chi(z)^2 \,\langle f_{\rm S} \rangle$, with the mean flux defined as:
\begin{eqnarray}
    \langle f_{\rm S} \rangle&=&\int_1^{3.5} \de\Gamma \int_{L_{\rm min}}^{L_{\rm max}} \de L_\gamma\,\Phi_{\rm S}(L_\gamma,z, \Gamma)\nonumber\\
    &\times& S\left(L_\gamma,z,\Gamma\right)(1-\Omega)
    \label{eq:win_astro}
\end{eqnarray}
The 3D power spectrum of the cross correlation between annihilating DM and blazars is given by:
\begin{eqnarray}
     &&P_{\rm BLA,DM}^{\rm 1h}(k,z) = \int_1^{3.5} \de\Gamma \int_{L_{\rm min}}^{L_{\rm max} }\de L_\gamma\,\frac{\Phi(L_\gamma,z,\Gamma)}{\langle f_{\rm S} \rangle} \nonumber\\
     &&~~\times S\left(L_\gamma,z,\Gamma\right)(1-\Omega)\, \frac{ \hat u_\kappa(k|M(L_\gamma,z)}{\Delta^2(z)}\\
     &&P_{\rm BLA,DM}^{\rm 2h}(k,z) = \int_{M_{\rm min}}^{M_{\rm max}} \de M\,\frac{\de n}{\de M} b_{\rm h} \frac{ \hat u_\kappa(k|M,z)}{\Delta^2(z)} \nonumber\\
     &&~~\times \int_1^{3.5} \de\Gamma \int_{L_{\rm min}}^{L_{\rm max}} \de L_\gamma\,b_{\rm S} \frac{\Phi(L_\gamma,z,{ \Gamma)}}{\langle f_{\rm S} \rangle} S\left(L_\gamma,z,{ \Gamma}\right)\,(1-\Omega) \nonumber \\
     &&~~\times P^{\rm lin}(k,z) \ ,
	\label{eq:PSastro}
\end{eqnarray}
where $b_{\rm S}$ is the bias of blazars with respect to the matter density, for which we adopt $b_{\rm S}(L_\gamma,z)=b_{\rm h}[M(L_\gamma,z)]$. The relation $M(L_\gamma,z)$ between the mass of the host halo and the luminosity of the hosted blazar is taken from \citep{Camera:2014rja}.

\subsection{Galactic Dark Matter modeling}
%-------------------------------------------------------------------

For the modeling of the signal expected from the Galactic subhalos, we followed the treatment of \cite{Ando:2009}. In general, the prediction of the APS for the Galactic component is the sum of two contributions, one arising from the main halo and the other originated by substructures. The main (smooth) halo contribution is subdominant for multipoles above a few: since we are dealing with multipoles larger than 50, it is here neglected. For the substructure contribution, we consider an anti-biased subhalo distribution, corresponding to the fiducial model A1 of \cite{Ando:2009}, with a boost factor of subhalos set to unity.

The subhalo number density as a function of the distance $f$ from the center of the Galaxy reads
\begin{eqnarray}
    n_{\rm sh}(r) &=& f \frac{M_{\rm vir,MW}}{2 \pi r_{-2}^3 M_{\rm min}} \gamma\left(\frac{3}{\alpha_E},\frac{2\,c_{-2}}{\alpha_E}\right)^{-1} \left(\frac{2}{\alpha_E} \right)^{3/\alpha_E -1} \nonumber \\
    &\times &\exp \left[-\frac{2}{\alpha_E} {\left(\frac{r}{r_{-2}} \right)^{\alpha_E}}  \right] \, ,
                \label{eq:nsh}
\end{eqnarray}
where $M_{\rm vir,MW}$ is the Milky Way virial mass, ${\alpha_E=0.68}$, $\gamma$ is the lower incomplete gamma function, ${r_{-2} = 0.81\,r_{200, \rm MW}}$, ${c_{-2}=r_{\rm vir}/r_{-2}}$ and the fraction $f$ of DM enclosed in subhalos is fixed to 0.2. The minimal subhalo mass is set at $M_{\rm min} = 10^{-6} M_\odot$. We do not include a truncation for the subhalo distribution at large radii. The angle-average number density referred to our position in the Galaxy is
\begin{equation}
    \overline{n}_{\rm sh} = \frac{1}{2} \int_{-1}^1 \diff \, {\cos} (\psi) \, \, n_{\rm sh} \, \left(r \left( \cos \psi \right) \right) \, ,
\end{equation}
where $r\left({\rm cos} (\psi)\right) = \sqrt{r_0^2 + s^2 - 2 \, r_\oplus \, s \, {\rm cos} (\psi)}$, $r_\oplus = 8.5$ kpc is our distance from the center of the Galaxy, $s$ represents the distance along the line of sight, and $\psi$ is the angle between the direction $\hat{n}$ of observation and the direction to the Galactic center.

Numerical simulations suggest that the mass distribution of subhalos follows a power-law behavior with the mass $M$ of the subhalo and can be written as
\begin{equation}
    \frac{\diff \overline{n}_{\rm sh}}{\diff M} = \overline{n}_{\rm sh} \left(r \right) \frac{\alpha_0 -1}{M_{\rm min}} \left(\frac{M}{M_{\rm min}} \right)^{-\alpha_0} \, ,
\end{equation}
where $\alpha_0=1.9$.

The subhalo luminosity $L$ for a subhalo with a mass $M$ depends on the particle properties of DM, $\langle \sv \rangle$ and $m_{\rm DM}$,  as well as on the energy spectrum $\diff N_{\rm ann}/\diff E$ associated to the channel under study. It reads
\begin{equation}
    L = \frac{\langle \sv \rangle M^2}{24 \pi m_\chi^2 r_s^3} \int \de E \, \frac{\de N_{\rm ann}}{\de E} \, .
\end{equation}
With the above ingredients, the APS for the Galactic subhalo contribution can then be expressed as
\begin{eqnarray}
    C_\ell^{ij} &=& \frac{1}{16 \pi^2 f_{sky}} \int_{M_{\rm min}}^{M_{\rm max}} \de M \int_{s_{\rm min}}^{s_{\rm max}} \de s \frac{L^i(M) L^j(M)}{s^2} \nonumber \\
    &\times& \frac{\de \overline{n}_{\rm sh}}{\de M}\, \tilde{u}^2_{\rm sh} \left(\frac{\ell}{s},M \right)\, ,
\end{eqnarray}
where $i$ and $j$ refer to energy bins. The integral along the line of sight is performed up to $s_{\rm max} = r_{\rm vir, MW}= 258$ kpc, and starts at $s_{\rm min} = \sqrt{L/ (4 \pi S_{\rm thr})}$.
In this specific case, we use a simplified approach: Instead of the full, \ie, $\Gamma$-dependent flux threshold, we use a fixed threshold averaged over $\Gamma$. More specifically, the thresholds (corresponding to the flux from 1 to 100 GeV) are set to $S_{\rm thr} = 10^{-10}$~cm$^{-2}$~s$^{-1}$ for the first ten energy bins (4GFL threshold) and $2 \times 10^{-10}$~cm$^{-2}$~s$^{-1}$ for two highest energy bins (3FHL threshold).
The value of $s_{\rm min}$ is thus chosen such that only unresolved Galactic subhalos are considered in the determination of the APS. We adopted an NFW profile for the internal density distribution of the subhalos and $\tilde{u}_{\rm sh}$ denotes its Fourier transform. The maximal halo mass for subhalos in our Galaxy, $M_{\rm max} = 10^{10} M_\odot$.\\

Summarizing, the APS involving DM is given by the sum of four terms, three extragalactic (the auto-correlation from extragalactic DM halos and the cross-correlation with BLLs and FSRQs) and the Galactic one, which is not expected to cross-correlate with extragalactic source populations.
Since the DM contributions are not flat in multipole but $\ell$-dependent, they are averaged in the multipole range considered for the determination of the \cp\ measurement in \citep{Ackermann:2018wlo}.

\subsection{Statistical framework}
%-------------------------------------------------------------------

We derive bounds on the DM annihilation cross section as a function of the DM mass by marginalizing over the uncertainties in the astrophysical \textit{background} model. This means that we are not using the approximation of a fixed background model obtained from the best-fit of the blazars-only case, on top of which the DM contribution is added. 

The naive approach to consider the full uncertainty would be to perform an extended parameter scan which would include at the same time the blazar parameters and the DM parameters. However, this is computationally very expensive. We instead use a method called importance sampling in order to recycle the information from background-only fits which significantly speeds up the calculation. The same approach has recently been used in a different context \citep{Kahlhoefer:2021sha}.

One \textit{side-product} of the \textsc{MultiNest} scan is a set of parameter vectors that follows the multidimensional posterior distribution. This set is provided in the so-called \emph{equal-weights} sample. We will apply importance sampling to obtain the posterior distribution of the full parameter space (blazars and DM) by using the equal-weights set of the fit to only blazars.

%=====================
%    \                                           |
%      \                                         |
%        \                                       |
\begin{figure*}[t]
  %\centering
    \includegraphics[width=.5\linewidth]{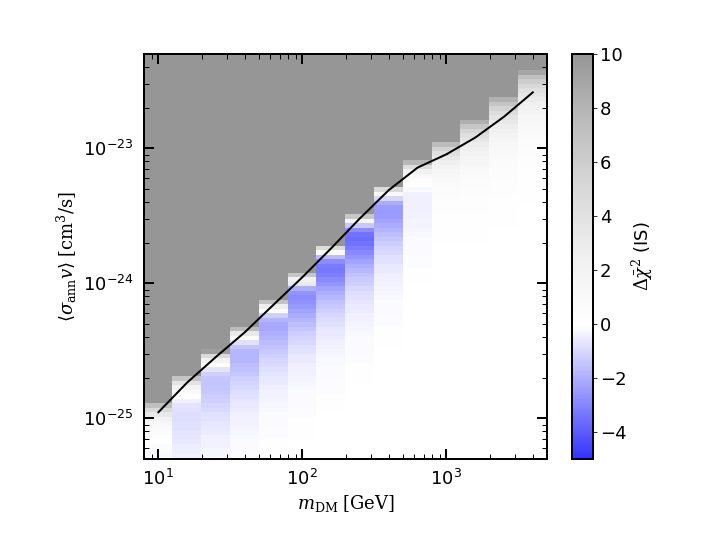}\includegraphics[width=0.5\linewidth]{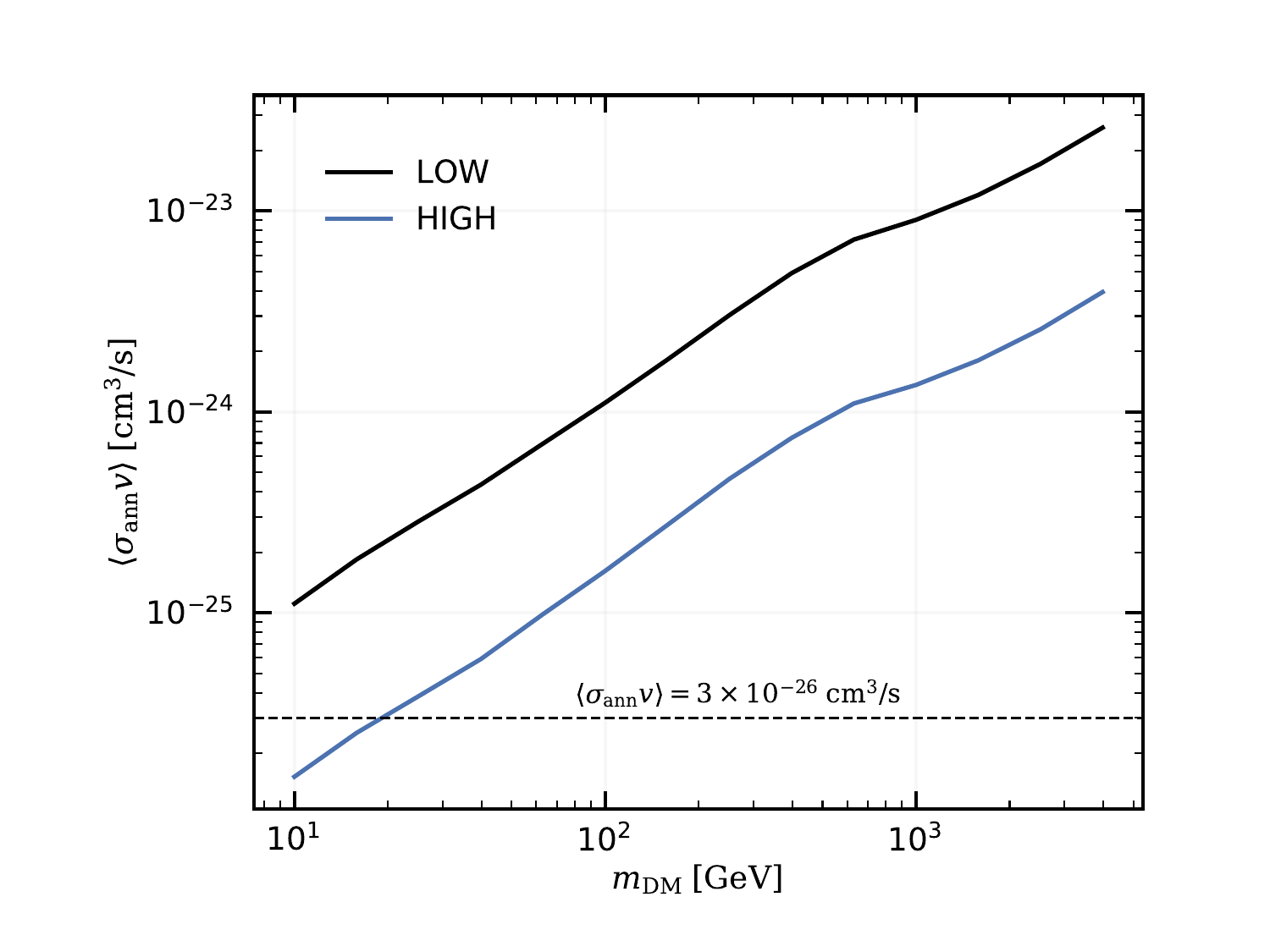}
    \caption{  
              95\% C.L. bounds on the annihilation cross section as a function of the DM mass, for annihilation into $\bar b b$.
              Colors in the left panel stand for the value of the marginalized $\Delta \chi^2$ derived from equation~\eqref{eq:IS_delta_chi2}. The left panel shows the constraints in the ``LOW'' scenario, while the right panel reports the comparison between ``LOW'' and ``HIGH'' cases.
            }
    \label{fig::DM_limit}
\end{figure*}
%                                      \         |
%                                        \       |
%                                          \     |
%=====================

First, we note that we can approximate the integral of the product of background, \emph{i.e.} the blazar, posterior and an arbitrary function $f$ over the blazar parameters, $\boldsymbol \theta_\mr{blz}$, by a sum over the parameter points in the set of the equal-weights sample: 
\begin{eqnarray}
    \label{eq:IS_sum}
    \int \diff \boldsymbol \theta_\mr{blz} && \frac{  {\mathcal{L}_0}  (\boldsymbol{\theta}_\mr{blz})
    \, p_0(\boldsymbol{\theta}_\mr{blz}) }{Z_0} \, f(\boldsymbol{\theta}_\mr{blz}) \\ \nonumber
    && \approx \frac{1}{N}\sum\limits_{i=1}^{N} f(\boldsymbol{\theta}_{\mr{blz},i}),
\end{eqnarray}
where $\mathcal{L}_0$ is the likelihood, $Z_0$ is the evidence,  and $p_0$ is the prior. The subscript $0$ indicates quantities that refer to the fit without DM. We note that the factor $ {\mathcal{L}_0}  (\boldsymbol{\theta}_\mr{blz})\,p_0(\boldsymbol{\theta}_\mr{blz}) / Z_0$ is by definition the posterior distribution. Furthermore, we know that the integral over the prior $p_0(\boldsymbol{\theta}_\mr{blz})$ is normalized to 1. 
If we set $f(\boldsymbol{\theta}_\mr{blz}) = Z_0/{\mathcal{L}_0}  (\boldsymbol{\theta}_\mr{blz})$ we see that the evidence is given by
\begin{eqnarray}
    \label{eq:IS_evidence}
    Z_0 =\frac{N}{\sum_{i=1}^{N} 1/\mathcal{L}_0(\boldsymbol{\theta}_{\mr{blz},i})} \, .
\end{eqnarray}

We can obtain the marginalized likelihood by integrating over the background parameters:
\begin{eqnarray}
    \label{eq:IS_marginal}
    \bar{\mathcal{L}} (\boldsymbol{\theta}_\mr{DM}) =
    \int d \boldsymbol \theta_\mr{blz} \; {\mathcal{L}} (\boldsymbol{\theta}_\mr{blz}, \boldsymbol{\theta}_\mr{DM})
    \, p(\boldsymbol{\theta}_\mr{blz}).
\end{eqnarray}
Here  $\boldsymbol \theta_\mr{DM} = \lbrace m_\mr{DM}, \langle \sigma_\mathrm{ann} v \rangle \rbrace$ denotes the DM parameters, ${\mathcal{L}} (\boldsymbol{\theta}_\mr{blz}, \boldsymbol{\theta}_\mr{DM})$ is the likelihood of the full parameter space, and $p(\boldsymbol{\theta}_\mr{blz})$ is the prior of the blazar parameters. Using equation~\eqref{eq:IS_sum} and assuming the same prior ($p(\boldsymbol{\theta}_\mr{blz}) = p_0(\boldsymbol{\theta}_\mr{blz})$), we see that the integral of equation~\eqref{eq:IS_marginal} is approximately given by the sums
\begin{eqnarray}
    \label{eq:IS_marginal_2}
    \bar{\mathcal{L}} (\boldsymbol{\theta}_\mr{DM}) \approx 
    \frac{\sum_{i=1}^{N} \frac{{\mathcal{L}} (\boldsymbol{\theta}_{\mr{blz},i}, \boldsymbol{\theta}_{\mr{DM}})}{{\mathcal{L}_0}  (\boldsymbol{\theta}_{\mr{blz},i})}}{\sum_{i=1}^{N} \frac{1}{{\mathcal{L}_0}  (\boldsymbol{\theta}_{\mr{blz},i})}}. 
 \end{eqnarray}
 In the next step, we can turn this equation into an expression for a marginalized $\chi^2$ by using the definition of our likelihood ($\mathcal{L} = \exp( -\chi^2/2 )$)
 \begin{eqnarray}
    \label{eq:IS_delta_chi2}
    \Delta&\bar{\chi}^2& (\boldsymbol{\theta}_\mr{DM}) =
    \bar{\chi}^2 (\boldsymbol{\theta}_\mr{DM}) -\bar{\chi}^2_0 \\ \nonumber
    &=&-2\log 
    \frac{\sum_{i=1}^{N} \exp \left( - \frac{ \chi^2 (\boldsymbol{\theta}_{\mr{blz},i}, \boldsymbol{\theta}_\mr{DM})-\chi^2_0(\boldsymbol{\theta}_{\mr{blz},i})}{2}\right)}{N}.
 \end{eqnarray}
Finally, we obtain the DM limit at the 95\% C.L. from the requirement $\Delta\bar{\chi}^2 (\boldsymbol{\theta}_\mr{DM}) \leq 3.84$.

In practice, we evaluate equation~\eqref{eq:IS_delta_chi2} on a grid of $m_\mr{DM}$ with 14 grid points logarithmically spaced between 10~GeV and 4~TeV. We note that the annihilation cross section $\langle \sv\rangle$ only changes the normalization of the \cp\ contributions but not the shape. The DM$\times$DM contributions scales with $\langle \sv\rangle^2$ while the DM$\times$BLZ contributions scales linearly with $\langle \sv\rangle$. So, we can tabulate the \cp\  contributions of background and DM at a reference value of $\langle \sv\rangle = 3\times10^{-26}\;\mr{cm^3/s}$. The equal-weights set contains $\mathcal{O}(10^5)$ parameter vectors. After the tabulation, the evaluation of equation~\eqref{eq:IS_delta_chi2} takes about one second. The importance sampling reduces the computing time by about a factor of ten since the typical number of evaluations required for a full parameter scan of blazar and DM parameters requires $~\mathcal{O}(10^6)$ evaluations. A further advantage of the importance sampling is that the tabulation can be parallelized to an arbitrary degree which is not possible for a Monte Carlo based parameter sampling with \textsc{MultiNest}.

%=====================
%    \                                           |
%      \                                         |
%        \                                       |
\begin{figure*}[t]
  %\centering
    \includegraphics[width=0.5\linewidth]{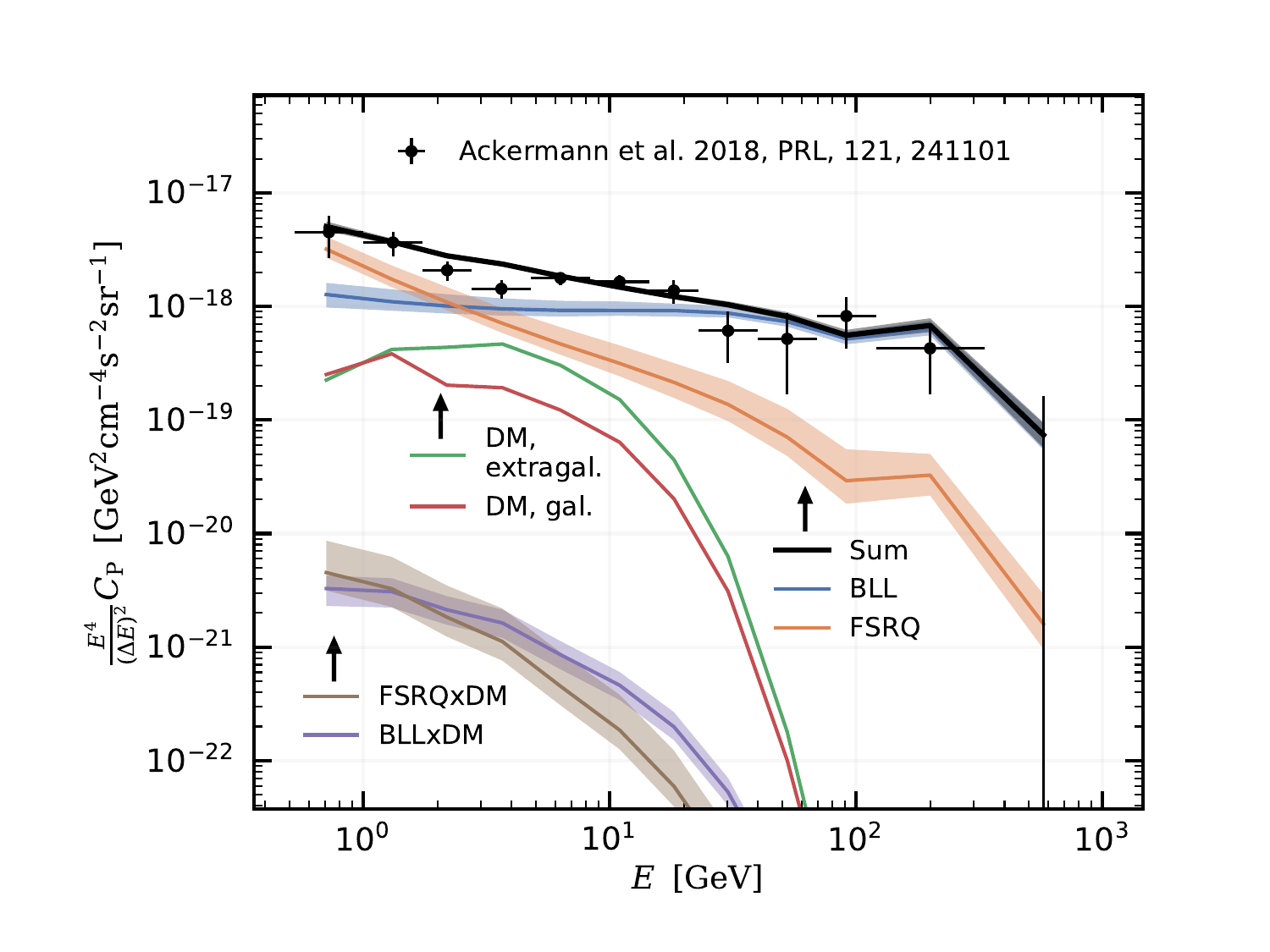}\includegraphics[width=0.5\linewidth]{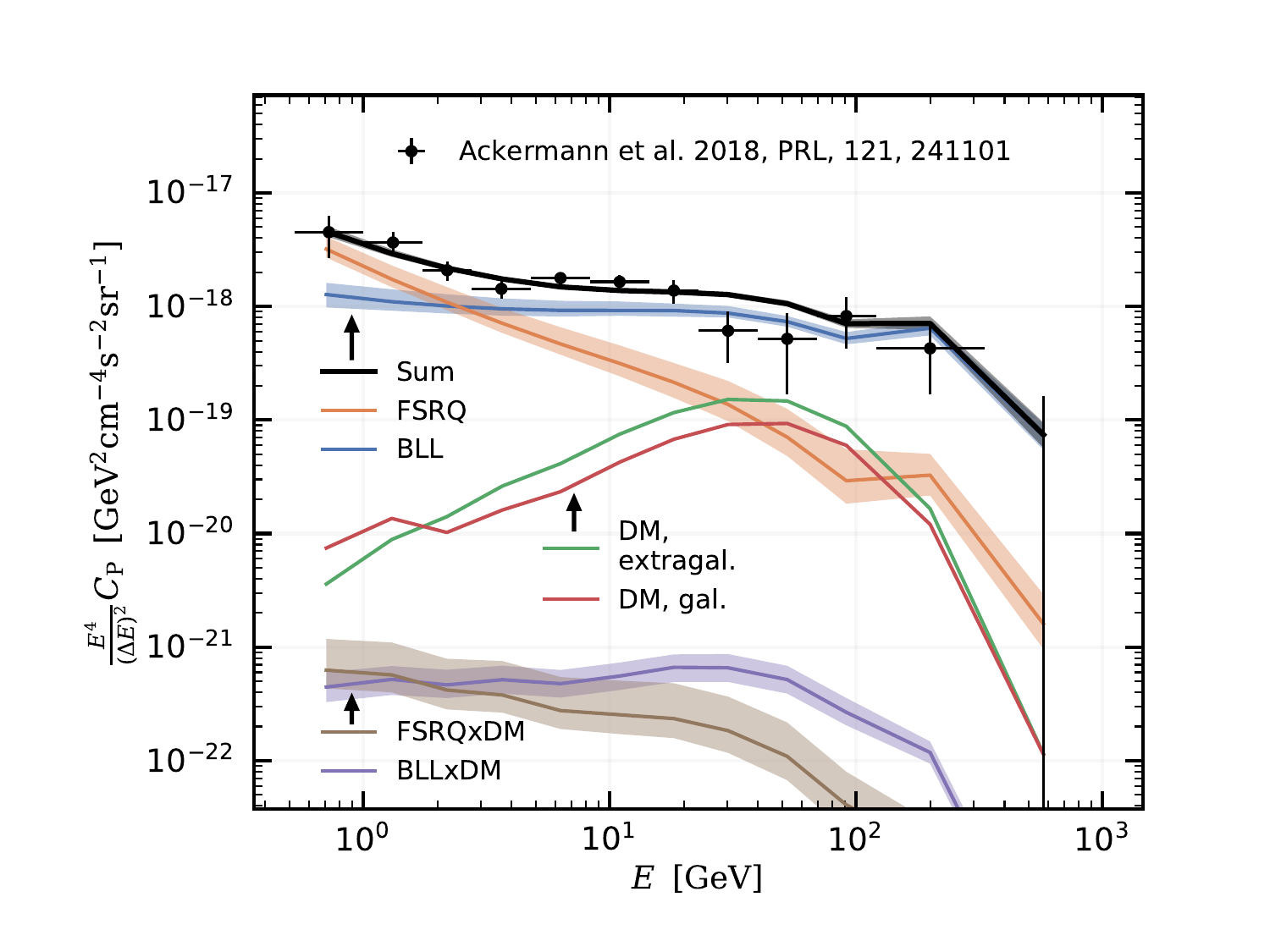}
    \caption{  
              Comparison of the \cp\ from blazars and DM. The DM contributions are shown at 100 GeV (left) and 1 TeV (right), and computed in the ``LOW'' scenario, for annihilation into $\bar b b$. In both panels we choose the value of $\langle\sigma_\mathrm{ann} v\rangle$ of DM at the limit derived in Figure~\ref{fig::DM_limit}. 
            }
    \label{fig::DM_at_limit}
\end{figure*}
%                                      \         |
%                                        \       |
%                                          \     |
%=====================

\subsection{Constraints on DM annihilation}
\label{subsec::DMcons}
%-------------------------------------------------------------------

We derive constraints on the annihilation of DM into a pair of $\bar b b$ quarks which serves as an illustrative example. The limits for other hadronic channels are expected at a similar level. The left panel of Figure~\ref{fig::DM_limit} shows the marginalized $\Delta \chi^2$ in the plane of DM mass and annihilation cross section as derived from equation~\eqref{eq:IS_delta_chi2}. For DM masses between 15 and 140 GeV a small DM contribution slightly improves the fit of the \cp\ data. However, it is statistically not significant. The maximal improvement of the $\Delta \bar{\chi}^2$ is $\sim 3.5$ which corresponds to a local significance of less than $2\sigma$ and an even smaller global significance. Consequently, we can derive DM limits as a function of the DM mass. In the fiducial setup, namely, using the ``LOW'' model for the concentration-mass relation of DM halos, see Section~\ref{sec::DM}, we can place an upper limit of $\langle \sv\rangle = 10^{-25}\;\mathrm{cm^3/s}$ on the annihilation cross section at the DM mass of 10 GeV. The limit gradually weakens to $\langle \sv\rangle = 3 \times 10^{-23}\;\mathrm{cm^3/s}$ at 4 TeV. In a more, aggressive setting for the concentration parameter, i.e., the ``HIGH'' model, we obtain a DM limit which is almost one order of magnitude stronger. The plot in the left panel of Figure~\ref{fig::DM_limit} shows the limits for the ``LOW'' model, while the comparison between the two cases is shown in the right panel. In the ``HIGH'' scenario, we can exclude a thermal WIMP for $m_\mr{DM} < 20$~GeV. 

As explained above the measurement of the APS is dominated by the Poisson noise term. For this reason, the DM bounds in Figure~\ref{fig::DM_limit} are weaker than the ones from other probes of the UGRB~\citep{DiMauro:2015tfa,2016PhR...636....1C}, such as the total intensity energy spectrum and the cross-correlation APS with gravitational tracers, which are less affected by the noise being linear instead of quadratic probes of the UGRB (see, e.g., Figure~4 in \cite{Regis:2015zka}). There are also other strategies of indirect DM searches using gamma-rays, \emph{e.g.} from the dwarf spheroidal galaxies or the Galactic Center, or using cosmic-ray antiprotons. Those DM limits are typically stronger by 1-2 orders of magnitude (see \emph{e.g.} \cite{Leane:2020liq,Slatyer:2021qgc} and references therein), however, those analyses are affected by different systematic uncertainties.

In Figure~\ref{fig::DM_at_limit}, we show the contribution of all the different components to the \cp\ for two exemplary DM masses of 100 GeV (left panel) and 1 TeV (right panel), in the ``LOW'' scenario. The DM components are evaluated at the $\langle\sigma_{\rm ann} v\rangle$ values corresponding to the limit shown in Figure~\ref{fig::DM_limit} for that DM mass. As expected, the blazar components are dominant and DM only provides a sub-dominant part. The largest DM contribution stems from the extragalactic DM halos, closely followed by the contribution of Galactic DM subhalos which is roughly smaller by a factor of 2, while the contribution of BLZ$\times$DM is smaller by 1-2 orders of magnitude. The uncertainty bands in Figure~\ref{fig::DM_at_limit} represent the 1$\sigma$ uncertainty in the blazar background models. In the ``HIGH'' scenario, the picture is similar but with the DM contribution strongly dominated by the extragalactic term.

Finally, we stress again that the absence of a DM signal in the \cp\ is a further confirmation that the sum of FSRQs and BLLs fully explains the entire UGRB anisotropy and no additional, weak component is required to match the data.

\newpage

%===================================================================
\section{Conclusions}
\label{sec::concl}
%===================================================================

In this work, we compared models of the GLF and SED of blazars to the latest measurement of the energy spectrum of the UGRB anisotropies \citep{Ackermann:2018wlo} and the properties of the resolved gamma-ray sources of the \Fermi-LAT 4FGL catalog. We considered two different blazar populations, distinguishing between BL Lacs (BLLs) and flat spectrum radio quasars (FSRQs). We found that BLLs and FSRQs can account for the totality of the UGRB anisotropy, with BLLs dominating the APS at high energies, while FSRQs being important at GeV energies. The derived models well reproduce the size and spectral features observed by \citep{Ackermann:2018wlo}, and the properties of source number counts of the 4FGL catalog. Our analysis significantly constrains the redshift and luminosity dependence of the blazar GLF and the spectrum of the SED in the unresolved regime. We also calculate the contribution of the unresolved population of FSRQs and BLLs to the the UGRB intensity spectrum finding a non-negligible contribution of about 30\% between 10 and 100 GeV and about 20\% at 1 GeV.

In the second part of the paper, we included a contribution to the UGRB arising from annihilating DM and performed a global fit to derive constraints on the particle DM parameters.
We computed both Galactic and extragalactic DM contributions, and included cross-terms in the APS, due to the cross-correlation of blazars with the DM halos hosting them. The dominant term arises from extragalactic DM halos, which strongly depends on the poorly-known description of DM subhalos. To bracket the uncertainty, we considered two different scenarios, ``LOW'' and ``HIGH'', which lead to an upper limit of $\langle \sv\rangle = 10^{-25}\;\mathrm{cm^3/s}$ and $\langle \sv\rangle = 1.5\times 10^{-26}\;\mathrm{cm^3/s}$, respectively, on the annihilation cross section at the DM mass of 10 GeV, for annihilation into bottom quarks.

The present analysis of the UGRB anisotropies is based on the measurement of \citep{Ackermann:2018wlo}, where no evidence for an $\ell$-dependent APS was found. Further data, and more resolved sources, would allow to reduce the level of the Poisson noise \cp\ and to measure an APS unveiling the large-scale clustering of gamma-ray sources. This would allow us to deepen our understanding of blazar populations, as well as exploit the APS observable in a much more powerful way in the context of DM bounds.

\medskip

%===================================================================
\section*{Acknowledgments}
%===================================================================
We thank S. Manconi for fruitful discussions throughout the work. 
Furthermore, we 
thank M. Di Mauro, 
T. Linden, and 
S. Manconi 
for helpful comments as well as a carefully reading of the manuscript.
This work  is supported by:  `Departments of Excellence 2018-2022' grant awarded by the Italian Ministry of Education, University and Research (\textsc{miur}) L.\ 232/2016; Research grant `The Anisotropic Dark
Universe' No.\ CSTO161409, funded by Compagnia di Sanpaolo and University of Turin; Research grant TAsP (Theoretical Astroparticle Physics) funded \textsc{infn}; Research grant `The Dark Universe: A Synergic Multimessenger Approach' No.\ 2017X7X85K funded by \textsc{miur};  Research grant ``Deciphering the high-energy sky via cross correlation'' and ``Unveiling Dark Matter and missing baryons in the high-energy sky'' funded by the agreement ASI-INAF n. 2017-14-H.0; Research grant ``From  Darklight  to  Dark  Matter: understanding the galaxy/matter connection to measure the Universe'' No. 20179P3PKJ funded by MIUR; MN: the material is based upon work supported by NASA under award number 80GSFC21M0002. 
M.K. is partially supported by the Swedish National Space Agency under contract 117/19 and the European Research Council under grant 742104.
E.P. is  supported by the Fermi Research Alliance, LLC under Contract No. DE-AC02-07CH11359 with the U.S. Department of Energy, Office of High Energy Physics.
Computations were enabled by resources provided by the Swedish National Infrastructure for Computing (SNIC) under the project No. 2020/5-463 partially funded by the Swedish Research Council through grant agreement no. 2018-05973. Furthermore, some computations in this work were performed with computing resources granted by RWTH Aachen University under the project No. rwth0578.

\bibliography{bibliography}{}

%===================================================================
%===================================================================

\clearpage
\newpage

\appendix

%===================================================================
\section{One or two populations: a phenomenological interpretation}
\label{sec::pheno}
%===================================================================

In this paper, we have considered two blazar populations, BLLs and FSRQs. Both populations are important and their sum provides a good fit to angular correlations of the UGRB. In more detail, FSRQs give the largest contributions to the \cp\ below a few GeV, while BLLs dominate at higher energies. As a consequence, we found that the FSRQs are responsible for the softening of the \cp\ at low energies, which is an interesting result because this softening has also been interpreted as a possible hint for a new source population \citep{Ando:2017alx}. In a similar spirit, also \citep{Ackermann:2018wlo} claimed that the \cp\ data indicates a hint for two populations. On the other hand, in \citep{Manconi:2019ynl} a good fit of the \cp\ data is obtained with one population using a model which combines FSRQs and BLLs into a single GLF and SED model.

Here, we explore the issue of one versus two populations more thoroughly by employing a \emph{phenomenological model} for the SED and GLF of two hypothetical source populations. The phenomenological model is slightly simplified as compared to the physical models for the GLFs and SEDs of the BLL and FSRQ source populations used in the main text. The most important difference is that the phenomenological model does not include a redshift dependence. However, the \cp\ data is not sensitive to redshift information because it is integrated along the line of sight. We find that our model is sufficient to explore the question about the number of source populations. Furthermore, it has the clear advantage that the large parts of the computations can be done analytically. We find that the width of the spectral index distribution is a key parameter to distinguish the scenarios of one or two populations.

\subsection{Definition of the GLF and SED}
%-------------------------------------------------------------------

The source count distribution as a function of the photon flux $S$ is modeled as a power law:
\begin{eqnarray}
    \label{eqn::simple_model__dNdF}
    \frac{\diff N}{\diff S} = A\,\left( \frac{S}{S_0} \right)^{-\gamma}\,,
\end{eqnarray}
where $A$ is an overall normalization, $\gamma$ is the power-law index, and $S_0$ is a reference flux. 
In the following, $S_0$ is fixed to $1\times10^{-10}\;\mathrm{cm^2 s^{-1}}$ and the fluxes $S$ and $S_0$ always refer to the photon flux in the energy bin from 1~GeV to 100~GeV. 
To relate the flux $S$ to the flux $S_i$ in a different energy bin $i$ we need the SED, which we model as a power law with an exponential cutoff:
\begin{eqnarray}
    \label{eqn::simple_model__SED}
    \frac{\diff N}{\diff E} = K \, \left( \frac{E}{E_0} \right)^{-\Gamma} \,
                                   \exp\left(-\frac{E}{E_c}\right)\,.
\end{eqnarray}
Here $K$ is the normalization, $E_0$ a reference energy, $\Gamma$ the photon spectral index, and $E_c$ the energy of the exponential cutoff. The exponential cutoff allows mimicking the attenuation of gamma rays at high energies. By definition, the flux $S$ is given by $S=\int_{1{\rm\,GeV}}^{100{\rm\,GeV}} \diff E \, \diff N/\diff E$. So, the ratio of the fluxes $S_i$ and $S$ is given by:
\begin{eqnarray}
    \label{eqn::simple_model__flux_relation}
    s_i = \frac{S_i}{S}  
    =\frac{ \int_{E_{\mathrm{min},i}}^{E_{\mathrm{max},i}} \diff E \, E^{-\Gamma} \,\exp(-{E}/{E_c}) }
          { \int_{1{\rm\,GeV}       }^{100{\rm\,GeV}     } \diff E \, E^{-\Gamma} \,\exp(-{E}/{E_c}) }\,.
\end{eqnarray}
Finally, we allow for an intrinsic distribution of the photon spectral indices following a Gaussian  with mean $\mu$ and width $\sigma$:
\begin{eqnarray}
    \label{eqn::simple_model__dNdGamma}
    \frac{\diff N}{\diff \Gamma} = \frac{1}{\sqrt{2\pi}\sigma}
                                            \exp{\left( - \frac{(\mu-\Gamma)^2}{2\sigma^2} \right)}\,.
\end{eqnarray}

Then, the \cp\ of unresolved point sources is given by:
\begin{eqnarray}
    \label{eqn::Cp_simple}
    C_\mathrm{P}^{ij} = \int \diff S\, \diff \Gamma \, S_i S_j \frac{\diff^2 N}{\diff S\diff \Gamma}\, 
                        \big[ 1- \Omega(S,\Gamma)\big].
\end{eqnarray}
where $\Omega(S,\Gamma)$ is the detector efficiency to resolve point sources, which we model as a $\theta$-function at a $\Gamma$-dependent flux threshold, $S_\mr{thr}$ (see main text).
By considering photon spectral indices between 1 and 3, the \cp\ is then calculated as:
\begin{eqnarray}\
    \label{eqn::Cp_simple_explicit}
    C_\mathrm{P}^{ij} &=& \int\limits_{1}^{3} \diff \Gamma \int\limits_0^{S_\mathrm{thr}(\Gamma)} \diff S\, \frac{1}{\sqrt{2\pi}\sigma}
                                            \exp{\left( - \frac{(\mu-\Gamma)^2}{2\sigma^2} \right)} \nonumber\\ 
                      &&  \qquad\qquad\qquad\times S^2 \, 
                                              s_i s_j A\,\left( \frac{S}{S_0} \right)^{-\gamma} \nonumber \\
                    &=& \int\limits_{1}^{3} \diff \Gamma \frac{1}{\sqrt{2\pi}\sigma}
                                            \exp{\left( - \frac{(\mu-\Gamma)^2}{2\sigma^2} \right)} \nonumber\\ 
                      &&  \qquad\qquad\qquad\times  
                                              s_i s_j \frac{A \, S_\mathrm{thr}^3(\Gamma)}{3-\gamma} \,\left( \frac{S_\mathrm{thr}(\Gamma)}{S_0} \right)^{-\gamma}\,.
\end{eqnarray}
We note that all the energy-information of the \cp\ is encoded in the factor $s_i s_j$ such that to a first approximation the slope, $\gamma$, of the \dnds\ is degenerate with the overall normalization, $A$. So, we fix $\gamma$ to a benchmark value of $2.2$ in the following analysis.

%=====================
%    \                                           |
%      \                                         |
%        \                                       |
\begin{deluxetable*}{cclcclclcc}
\tablenum{3}
\caption{
    Fit results of the phenomenological model.
   \label{tab::PhenoModel}
    }
\tablewidth{700px}
\tablehead{
$                                                                      $& \multicolumn4c{w/o $\Gamma$ distribution}                                                && \multicolumn4c{w/ $\Gamma$ distribution}                                             \\
\cline{2-5}
\cline{7-10}
$                                                                      $& 1 pop.                   &\ & \multicolumn2c{2 pop.s}                              &\ \ \ & 1 pop.                   &\ & \multicolumn2c{2 pop.s}                                 \\
\cline{2-2}
\cline{4-5}
\cline{7-7}
\cline{9-10}
$                                                                      $&                            && pop. a                     & pop. b                        &&                            && pop. a                     & pop. b
}
\startdata
$\log_{10}\left({A}\;\mathrm{[cm^{-2}s^{-1}sr]}\right)                 $&$ 12.18_{-  0.16}^{+  0.17}$&&$ 11.87_{-  0.13}^{+  0.36}$&$ 11.88_{-  0.17}^{+  0.34}$   &&$ 12.26_{-  0.18}^{+  0.15}$&&$ 11.71_{-  0.20}^{+  0.81}$&$ 11.52_{-  0.42}^{+  0.96}$\\
${\mu}                                                                 $&$  2.11_{-  0.03}^{+  0.04}$&&$  1.86_{-  0.11}^{+  0.19}$&$  2.50_{-  0.29}^{+  0.18}$   &&$  2.18_{-  0.05}^{+  0.06}$&&$  2.02_{-  0.07}^{+  0.22}$&$  2.44_{-  0.32}^{+  0.17}$\\
${\sigma}                                                              $&                         -  &&                         -  &                         -     &&$  0.34_{-  0.08}^{+  0.07}$&&$  0.30_{-  0.09}^{+  0.13}$&$  0.30_{-  0.08}^{+  0.18}$\\
$\log_{10}\left({E_c}\;\mathrm{[GeV]}\right)                           $&$  2.19_{-  0.13}^{+  0.12}$&& \multicolumn2c{$  1.98_{-  0.15}^{+  0.12}$}               &&$  1.88_{-  0.12}^{+  0.10}$&& \multicolumn2c{$  1.88_{-  0.11}^{+  0.10}$}            \\
$k_{C_\mathrm{P}}                                                      $&$  1.00_{-  0.50}^{+  0.29}$&& \multicolumn2c{$1.00_{-  0.50}^{+  0.29}$}                 &&$  1.00_{-  0.24}^{+  0.96}$&& \multicolumn2c{$0.97_{-  0.42}^{+  0.38}$}              \\
$\chi^2                                                                $& 85.2                       && \multicolumn2c{74.9}                                       && 74.7                       && \multicolumn2c{74.5}                                    \\
$\Delta\chi^2                                                          $& \multicolumn4c{10.3}                                                                     && \multicolumn4c{0.2}                                                                   \\
\enddata
\end{deluxetable*}
%                                      \         |
%                                        \       |
%                                          \     |
%=====================

%=====================
%    \                                           |
%      \                                         |
%        \                                       |
\begin{figure*}[t!]
    \includegraphics[width=0.5\linewidth]{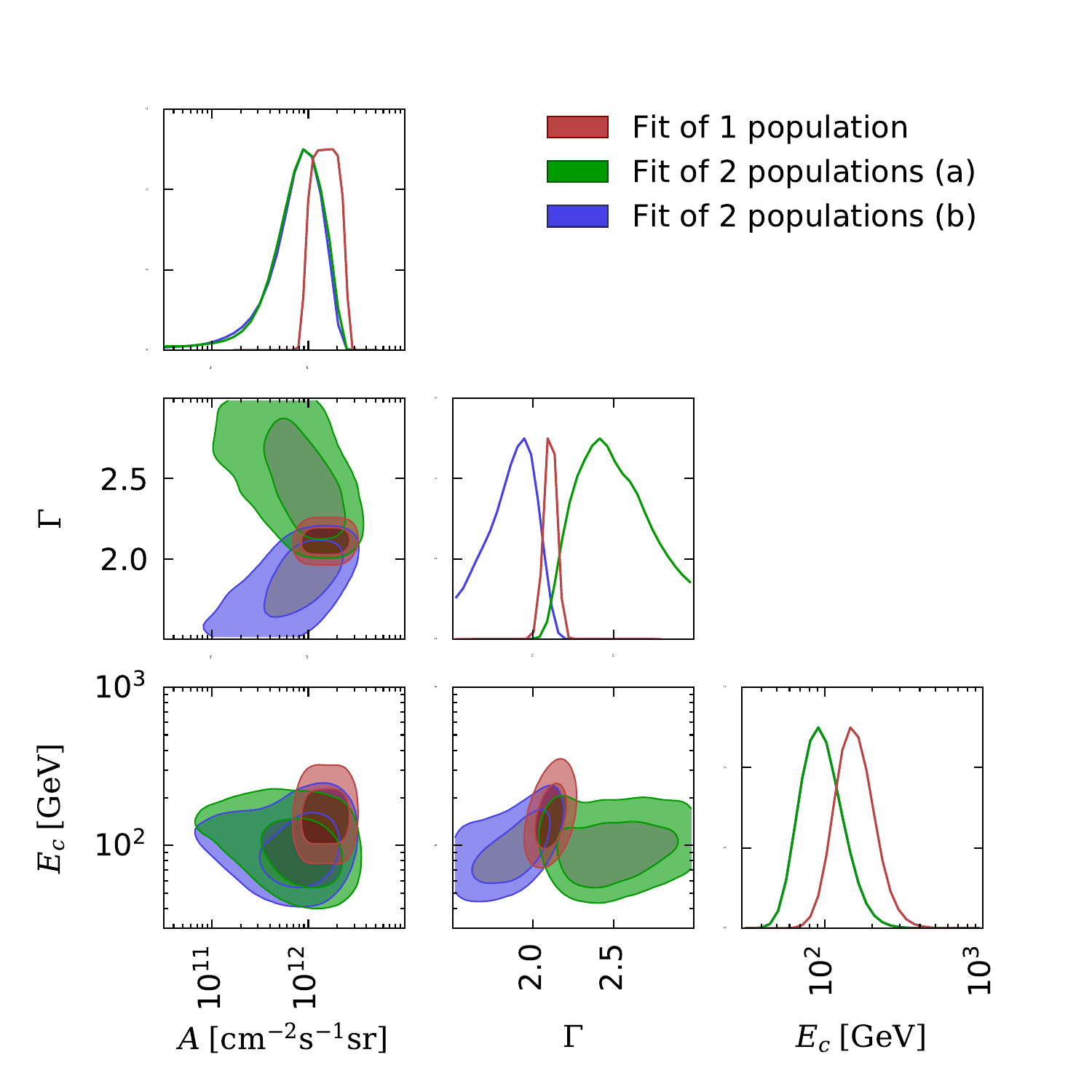}\includegraphics[width=0.5\linewidth]{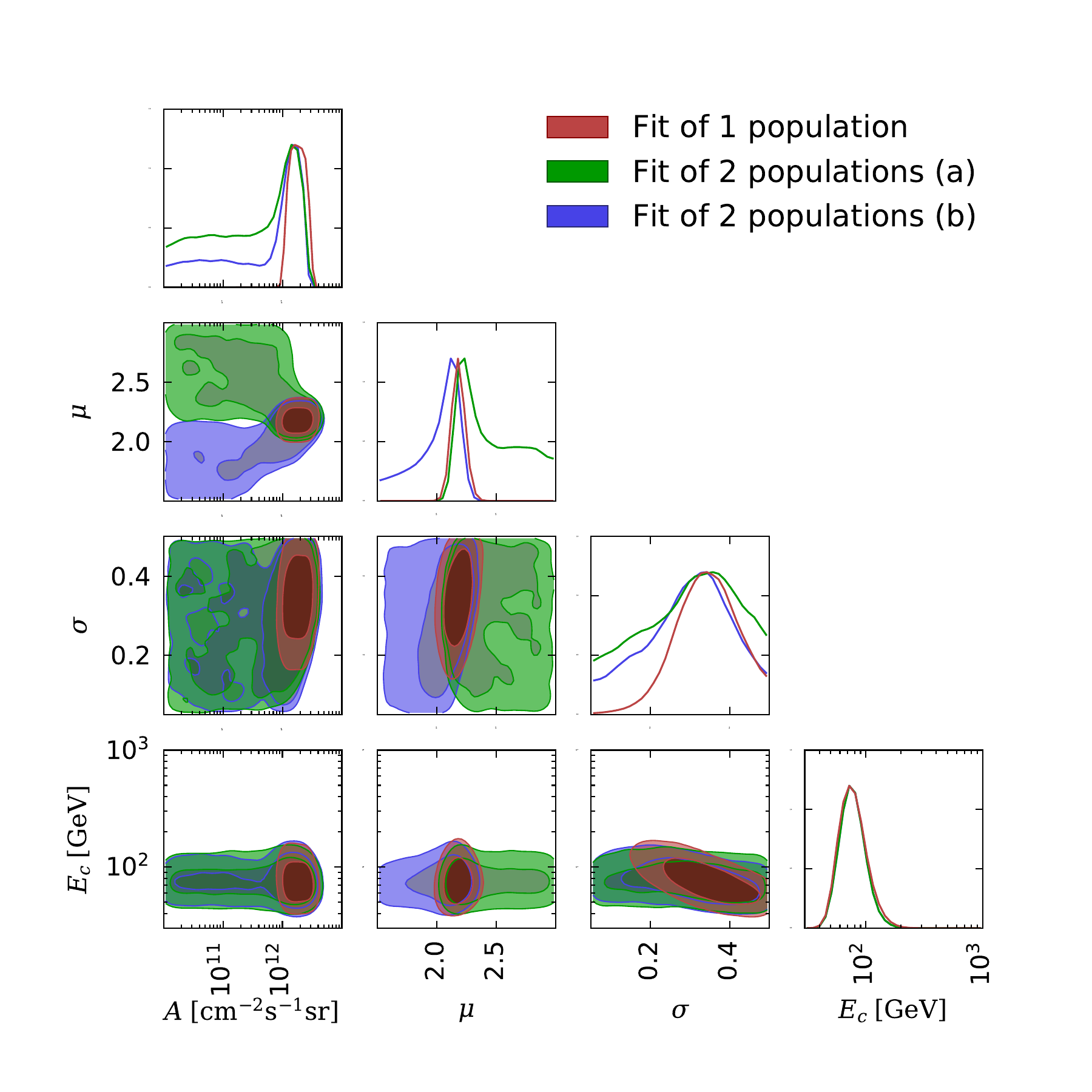}
    \caption{  
         	 Triangle plots showing the best fit region of the phenomenological model fitted to the \cp\ data for different fit setups.  On the diagonal, we display the marginalized likelihood for all parameter and the remaining panels below the diagonal contain the 1 and 2 $\sigma$ contours derived from the marginalized likelihood in two dimensions for each parameter combination.  The red contours and lines correspond to a fit with a single population, while the green and blue contours correspond to the two different populations of a fit containing two populations (labeled a and b). Left panel: Phenomenological model with a fixed photon spectral index. Right panel: Phenomenological model with a Gaussian distribution of the photon spectral index.
		     \label{fig::TrianglesPhenoModel}  
            }
\end{figure*}
%                                      \         |
%                                        \       |
%                                          \     |
%=====================

\subsection{Fits and results for the phenomenological model}
%-------------------------------------------------------------------

In this section, we perform a total of four fits. The fits differ by the number of source populations (one or two) and by the inclusion of a distribution in the photon spectral index.

%=====================
%    \                                           |
%      \                                         |
%        \                                       |
\begin{figure*}[t]
    \includegraphics[width=0.5\linewidth]{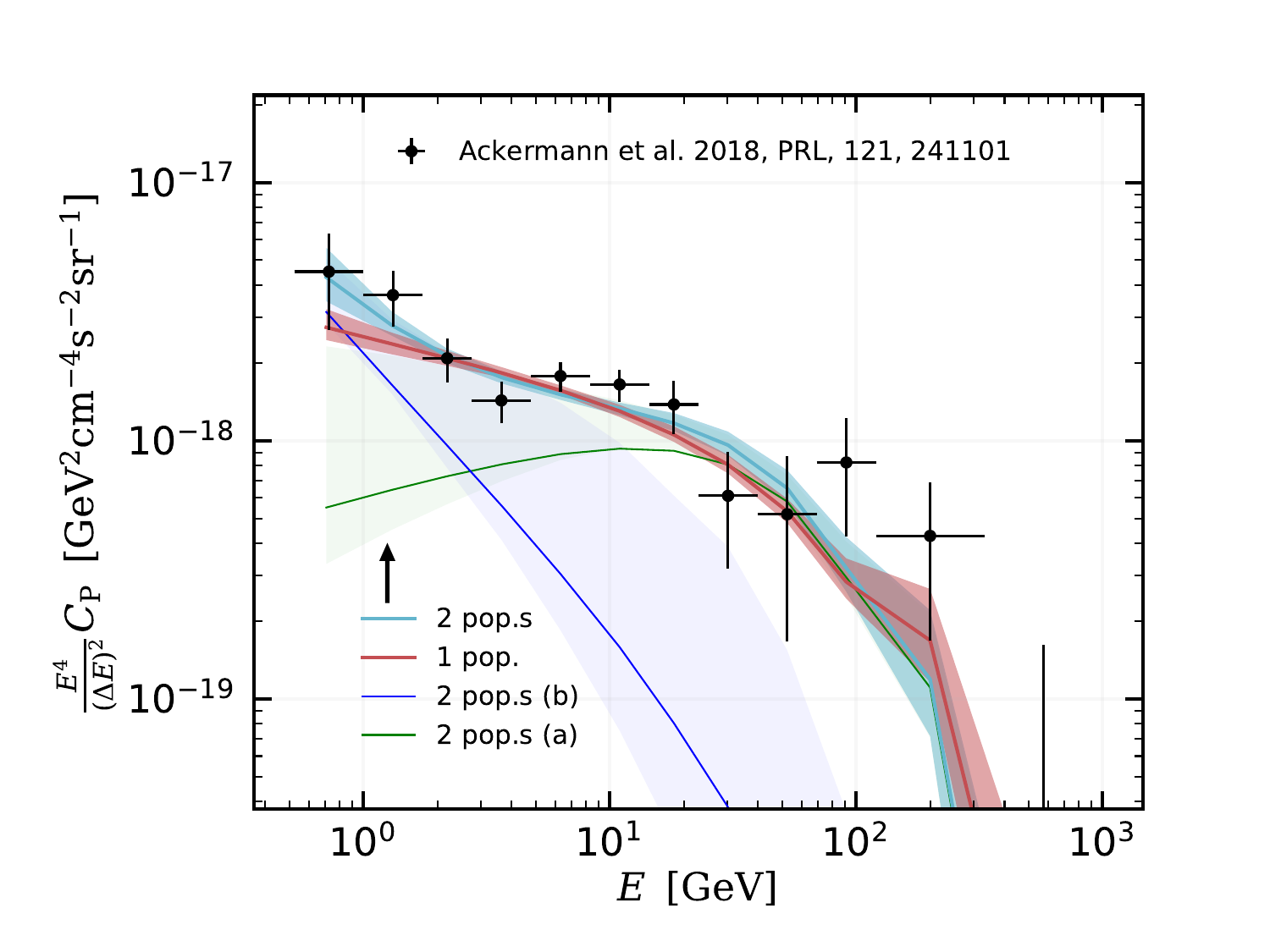}\includegraphics[width=0.5\linewidth]{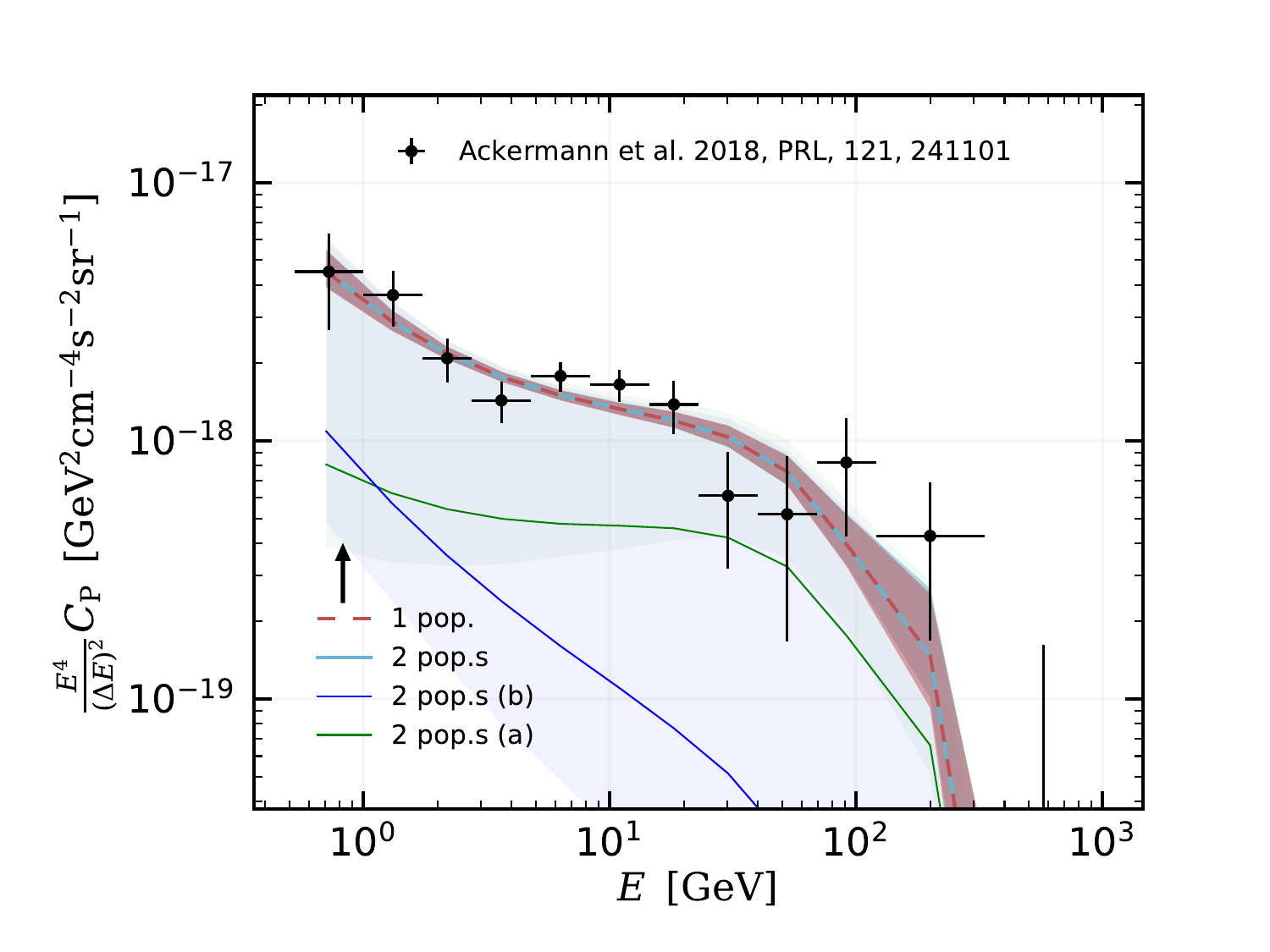}
    \caption{  
         	 Left panel: Phenomenological model with a fixed photon spectral index. Right panel: Phenomenological model with a Gaussian distribution of the photon spectral index.
		     \label{fig::Cp_spectrum_phenomodel}  
            }
\end{figure*}
%                                      \         |
%                                        \       |
%                                          \     |
%=====================

In the first two fits, we neglect the distribution of spectral indices, namely we force $\sigma=0$. Then, the first analysis employs a single source population with the SED and GLF specified in the previous paragraph. The free parameters are the normalization of the GLF ($A$), the photon spectral index ($\mu$), and the energy cutoff of the SED ($E_c$). Furthermore, we use a nuisance parameter ($k_{C_\mr{P}}$) that varies the value of flux threshold (see main text for more details). This fit has 4 free parameters.
In the second analysis, we use two source populations, labeled $a$ and $b$. They have the same functional form, but different parameters: in particular, for the normalization of the GLF and the photon spectral index. However, we force them to have the same value for the cutoff, $E_c$. Together with the nuisance, this amounts to 6 free parameters. To avoid an extra degeneracy in the fit we restrict the photon spectral indices to $\mu_a < \mu_b$.

The third and fourth fits are very similar to the first two, but we allow for a distribution in $\Gamma$. Consequently, this leads to one ($\sigma$) and two ($\sigma_a$ and $\sigma_b$) more parameters in the fits, respectively.
We use the \textsc{MultiNest} code to perform the four fits and to obtain the posterior distributions presented in the following. Uncertainties are stated in the Bayesian statistical framework. 

\bigskip

%=====================
%    \                                           |
%      \                                         |
%        \                                       |
\begin{figure}[b]
    \includegraphics[width=1.0\linewidth,trim={1cm 0cm 1cm 1cm},clip]{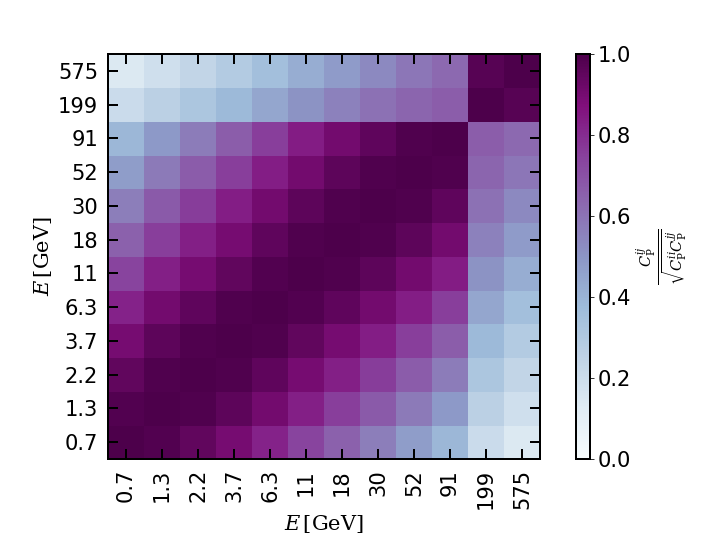}
    \caption{  
         	 \cp\ energy-correlation matrix for the phenomenological model in the setup with only one population but including a $\Gamma$-distribution. 
         	 \label{fig::Cp_Rij}  
            }
\end{figure}
%                                      \         |
%                                        \       |
%                                          \     |
%=====================

The results of our four fits are presented in Table~\ref{tab::PhenoModel} and in Figure~\ref{fig::TrianglesPhenoModel}. The $\chi^2$/dof of all four fits are close to 1 and we conclude that all of them give a good fit to the \cp\ data. However, when comparing the $\chi^2$s of the first and second fit, \emph{i.e.} the fits without the $\Gamma$ distribution, we see that including a second population reduces the total $\chi^2$ by 10.3, which formally corresponds to a statistical shift by $2.8\sigma$. So, this could be interpreted as a small hint for two populations. However, if we look at the results with a $\Gamma$ distribution the conclusion changes. Here the $\chi^2$ only decreases by 0.2 when a second population is introduced, which is not significant. So, we cannot distinguish between one or two populations based only on the \cp\ data. This can also be seen from the triangle plots in Figure~\ref{fig::TrianglesPhenoModel}. Without allowing for a $\Gamma$ distribution, the red posterior contours (one population) are not fully included in the contours of either the green or blue (population $a$ or $b$ of two population fit). Furthermore, both populations, $a$ and $b$ are required to have a relatively high normalization of $A>10^{-11}\,\mathrm{cm^{-2}s^{-1}sr^{-1}}$. On the other hand, if we allow for a $\Gamma$ distribution, the red contours are fully included in both the green and the blue contours. And, in the case of 2 populations, one of the normalizations $A$ can be pushed to negligible values of $10^{-12}\,\mathrm{cm^{-2}s^{-1}sr^{-1}}$. We compare the \cp\ energy autocorrelations of the best-fit to data in Figure~\ref{fig::Cp_spectrum_phenomodel}. It is clearly visible that, in the case without $\Gamma$ distribution, two populations provide a better fit, while in the case with the $\Gamma$ distribution there is almost no difference. 

Finally, we have a look at the \cp\ energy-correlation matrix (coefficients defined as $C_\mathrm{P}^{ij}/\sqrt{C_\mathrm{P}^{ii} C_\mathrm{P}^{jj} }$). By definition, the diagonal coefficients are equal to one. If there was only one source population with a single and fixed photon spectral index, also the off-diagonal coefficients would be equal to 1. Instead, the \cp\ measurement \citep{Ackermann:2018wlo} showed that the off-diagonal coefficients are $\sim 0.6$, which was interpreted as an indication for two populations. In Figure~\ref{fig::Cp_Rij}, we show that also a single source population with a distribution of photon spectral indices provides the correct off-diagonal pattern.

\subsection{Conclusions}
%-------------------------------------------------------------------

We conclude that, based on the \cp\ measurement itself, it is not possible to distinguish between two populations with narrow spectral index distributions and a single population with a broader distribution. In the latter case, the required value for the width of the $\Gamma$ distribution is $\sigma = 0.34_{-0.08}^{+0.07}$. Additional information can help to solve the riddle.
In the analysis of the main text, we included a physical model and the constraining power from the 4FGL catalog. 
This allowed us to break the degeneracy and to determine the presence of two populations, BLL and FSRQ.

%===================================================================
\section{Calculation of the Bayes factor}
\label{sec::bayes_factor_Cp}
%===================================================================

In the main text, we use a physical model with two population, BLL and FSRQ. Because of the 4FGL catalog prior, this model is significantly preferred over a model with only one BLL populations. This can be quantified by the calculation of Bayes factors, which is biefly summarized here. The Bayes factor is defined as the ratio of evidences between the two hypotheses: two populations (labeled all) and one population (labeled BLL). For flat priors, it becomes
\begin{eqnarray}
    \label{eq:BF_1}
    B = \frac{Z_{\rm all}}{Z_{\rm BLL}} = 
    \frac{ \int \diff \boldsymbol \theta_\mr{blz}  {\mathcal{L}_{\rm 4FGL}}  (\boldsymbol{\theta}_\mr{blz}) {\mathcal{L}_{C_{\rm P},{\rm all}}}  (\boldsymbol{\theta}_\mr{blz})}{ 
           \int \diff \boldsymbol \theta_\mr{blz}  {\mathcal{L}_{\rm 4FGL}}  (\boldsymbol{\theta}_\mr{blz}) {\mathcal{L}_{C_{\rm P},{\rm BLL}}}  (\boldsymbol{\theta}_\mr{BLL})} \,.\;\;\;\;  
\end{eqnarray}
Here, ${\mathcal{L}_{C_{\rm P},{\rm all}}}$ and ${\mathcal{L}_{C_{\rm P},{\rm BLL}}}$ are the likelihoods for the \cp\ data using the sum of BLLs plus FSRQs and only BLLs, respectively. We note that the \cp\ likelihoods also depend on the nuisance parameters $\boldsymbol{\theta}_n$ which we suppress in the notation.
In analogy to Eq.~\eqref{eq:IS_sum}, we can approximately replace the integrals by sums,  
\begin{eqnarray}
    \label{eq:BF_2}
    B \approx
    \frac{ \sum_i  {\mathcal{L}_{C_{\rm P},{\rm all}}}  (\boldsymbol{\theta}_{\mr{blz},i})}{ 
           \sum_i  {\mathcal{L}_{C_{\rm P},{\rm BLL}}}  (\boldsymbol{\theta}_{\mr{BLL},i})}\,,
\end{eqnarray}
where $i$ runs over the all points in the equal-weights sample of the 4FGL-only fit. Finally, we note one subtlety: The normalization of the \cp\ model strongly depends on the exact value of the catalog threshold which is only known approximately. To take the uncertainty due to the threshold into account, we profile over the \cp\ normalization, individually for each parameter point $i$. The mean renormalization factors are 1.04 and 1.16 for the hypotheses of two populations and one population, respectively.

%===================================================================
\section{Resolving previous UGRB anisotropy measurements}
\label{sec::prev}
%===================================================================

%=====================
%    \                                           |
%      \                                         |
%        \                                       |
\begin{figure}[t!]
    \includegraphics[width=1.0\linewidth,trim={1cm 0.9cm 2cm 1cm},clip]{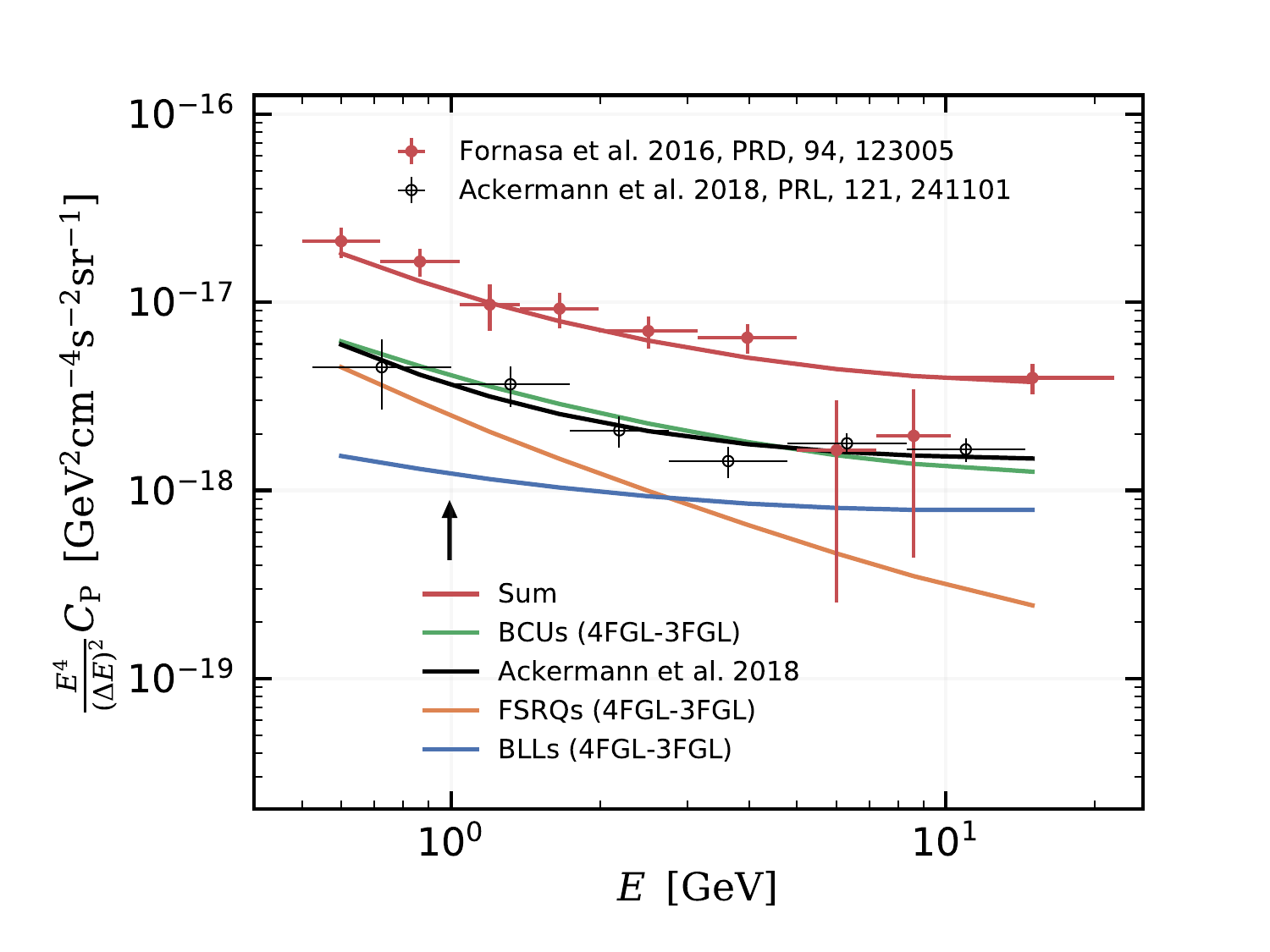}
    \caption{  
         	 Comparison between the measurements from \citep{Fornasa:2016ohl} (red points) and \citep{Ackermann:2018wlo} (black points). The \cp's of the populations of FSRQs, BLLs, and BCUs of the 4FGL not yet resolved in the 3FGL, are shown in orange, blue and green, respectively, and the best-fit to the \cp\ of \citep{Ackermann:2018wlo} is given by the black curve. Their sum is shown with the red curve, which well-reproduces the measurement of \citep{Fornasa:2016ohl}. The comparison is restricted to energies below 10 GeV, see text.
		     \label{fig::Cp_diagonal_3fgl_contributions}
            }
\end{figure}
%                                      \         |
%                                        \       |
%                                          \     |
%=====================

As mentioned in the main text, the UGRB emission is exposure-dependent: the more the LAT observes, the more sensitive is the survey to fainter sources, and consequently less unresolved emission contributes to the UGRB. We show this effect by comparing the UGRB anisotropy energy spectrum measured by \citep{Fornasa:2016ohl}, which masked sources from the  3FGL source catalog \citep{Acero:2015hja} based on 4 years of LAT survey, with the one in \citep{Ackermann:2018wlo}. In Figure~\ref{fig::Cp_diagonal_3fgl_contributions} we report both measurements in red and black, respectively. The difference between the two measurements, which consider similar total mission times (7.5 years in \citep{Fornasa:2016ohl} and 8 years in \citep{Ackermann:2018wlo}), is mainly due to the difference in the number of resolved sources masked away from the analyzed maps: the 4FGL counts approximately 2000 more sources than the 3FGL. We test this by evaluating the anisotropy energy spectrum of the populations of FSRQs, BLLs, and BCUs of the 4FGL not yet resolved (i.e., not present) in the 3FGL (let us call it $\Delta C_{\mathrm{P,4FGL-3FGL}}$). We verify that the sum of the best-fit model of the measurement in \citep{Ackermann:2018wlo} plus the $\Delta C_{\mathrm{P,4FGL-3FGL}}$ reproduces the anisotropy energy spectrum as measured by \citep{Fornasa:2016ohl}. In agreement with our expectations, the results in Figure~\ref{fig::Cp_diagonal_3fgl_contributions} show that there is a transition between FSRQs and BLLs in the $\Delta C_{\mathrm{P,4FGL-3FGL}}$ when going from lower to higher energies. So it is plausible to observe a similar transition in the \cp\ measurement of \citep{Ackermann:2018wlo}.

This test is performed considering only energies up to 10 GeV. The measurement by \citep{Ackermann:2018wlo} masks also 3FHL sources above those energies, making the comparison with \citep{Fornasa:2016ohl} less straightforward and beyond the scope of this study.

\end{document}